\documentclass[aps,twocolumn,floatfix]{revtex4-1}
\usepackage{bm, dcolumn, graphicx, amsmath, amssymb, amsfonts, dsfont}

\newcommand{\tbf}{\textbf}
\newcommand{\trm}{\textrm}
\newcommand{\mrm}{\mathrm}

\begin{document}
\title{Effect of trivial bands on chiral anomaly-induced longitudinal magnetoconductivity in Weyl semimetals}
\author{Jeonghyeon Suh$^{1}$}
\author{Hongki Min$^{1,2}$}
\email{hmin@snu.ac.kr}
\affiliation{$^{1}$ Department of Physics and Astronomy, Seoul National University, Seoul 08826, Korea}
\affiliation{$^{2}$ Center for Theoretical Physics, Seoul National University, Seoul 08826, Korea}
\date{\today}

\begin{abstract}
Including the effect of the trivial band near Weyl nodes, we evaluate the longitudinal magnetoconductivity (LMC) of Weyl semimetals along the magnetic field direction using the Boltzmann magnetotransport theory, and study its dependence on the magnetic field, Fermi energy, and temperature. We find that for weak internode and node-trivial band scatterings, the LMC is quadratic in the magnetic field and is inversely proportional to the fourth power of the Fermi energy at high densities due to internode scatterings, and to the square of the Fermi energy at low densities due to scatterings between a Weyl node and the trivial band. In the case of strong internode and node-trivial band scatterings, the magnetic field-driven anisotropy induced by the phase-space volume element and the orbital magnetic moment cannot be neglected. As a result, the LMC exhibits a significantly different trend compared to that in the weak internode and node-trivial band scattering limit. Finally, we calculate the temperature dependence of the LMC in the strong inelastic scattering limit and obtain its asymptotic behaviors at low and high temperatures, respectively, demonstrating that the temperature dependence is strongly affected by the existence of the trivial band.
\end{abstract}

\maketitle

\section{Introduction}
When an electric field $\bm{E}$ and magnetic field $\bm{B}$ are applied to Weyl semimetals (WSMs), the chiral anomaly \cite{Adler1969, Bell1969} breaks the number conservation of electrons in each Weyl node with chirality $\chi = \pm 1$, given by 
\begin{equation}
	\frac{\partial n_\chi}{\partial t} = \chi \frac{e^2 \bm{E} \cdot \bm{B}}{4\pi^2 \hbar^2 c},
\end{equation}
where $n_\chi$ is the electron number density at node $\chi$. It effectively pumps electrons from one node to another, leading to the charge imbalance between the nodes, which induces the positive longitudinal magnetoconductivity (LMC) along the magnetic field direction in WSMs \cite{Son2013,Kim2013,Burkov2014,Huang2015,Burkov2015,Spivak2016,Zyuzin2017,Zhang2016,Sharma2017,Andreev2018,Kundu2020,Shao2021}. Utilizing the relaxation time approximation $\frac{d f_\chi}{dt} = - \frac{f_\chi}{\tau^\trm{a}}$, where $f_\chi$ is the nonequilibrium distribution function and $\tau^\trm{a}$ is the anomalous relaxation time which characterizes the relaxation rate of the chiral charge imbalance, Ref.~\cite{Son2013} found that the LMC is given by
\begin{equation}
	\label{eq:LMC_approx}
	\delta \sigma (\bm{B}) = \frac{\mrm{g} e^2}{4\pi^2 \hbar c} \frac{v}{c} \frac{(eB)^2 v^2}{\varepsilon_\mrm{F}^2} \tau^\trm{a} (\varepsilon_\trm{F}),
\end{equation}
where $\mrm{g}$ is the number of Weyl node pairs, $v$ is the Fermi velocity of Weyl nodes, and $\varepsilon_\trm{F}$ is the Fermi energy measured from the Weyl point energy.

\begin{figure}[t]
	\includegraphics[width=0.7\linewidth]{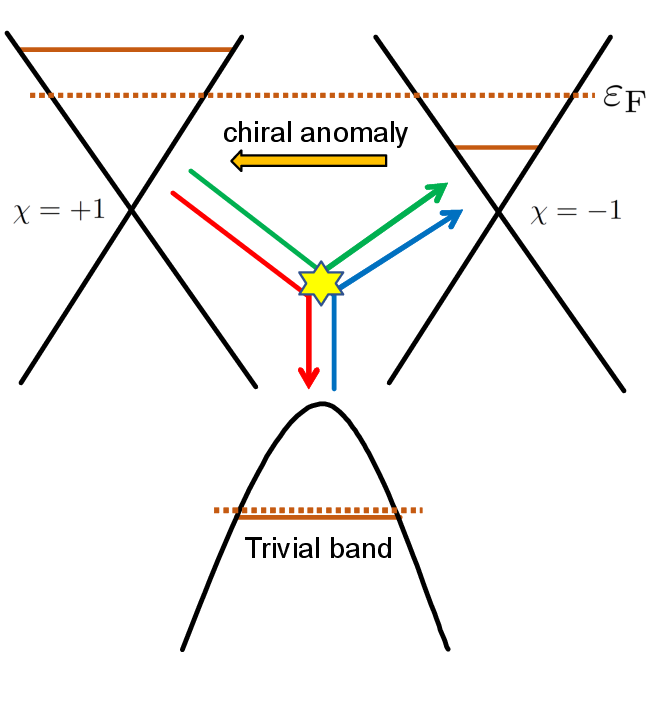}
	\caption{Schematic illustration of the charge imbalance caused by the chiral anomaly and the associated scattering processes leading to a steady state. The colored arrows represent scatterings by impurities indicated by the star at the center. The global and local Fermi levels of each node are represented by the dotted and solid brown lines, respectively. Here, $\bm{E} \cdot \bm{B} > 0$ is assumed.} 
	\label{fig:qualitative}
\end{figure}

Understanding the anomalous relaxation time is essential to properly estimate the LMC. Conventional theories have assumed that for a steady state, the charge pumping by the chiral anomaly is entirely compensated by the scatterings between the Weyl nodes. Therefore, $\tau^\trm{a}$ has been considered as the internode scattering time \cite{Son2013,Kim2013,Huang2015,Spivak2016,Zhang2016,Zyuzin2017,Kundu2020,Shao2021,Parameswaran2014,Sukhachov2022,Nishida2023}. However, as shown in Fig.~\ref{fig:qualitative}, scatterings between the Weyl nodes and the neighboring trivial band can also affect the chiral charge imbalance. For instance, for the $+$ node, the influx of electrons by the chiral anomaly is balanced by the outflux not only by internode scatterings (green arrow) but also by scatterings between the trivial band and the $+$ node (red arrow). In addition, the trivial band which is not involved in the charge pumping by the chiral anomaly, maintains a steady state by balancing the influx by scatterings between the trivial band and the $+$ node (red arrow) and the outflux by scatterings between the trivial band and the $-$ node (blue arrow). This trivial band effect becomes dominant near the Weyl point due to the low density of states (DOS) at the Weyl nodes, making scatterings between the Weyl nodes and the trivial band prevail over internode scatterings. On the contrary, as the Fermi energy increases, the effects of internode scatterings become significant, and the anomalous relaxation time would exhibit distinct dependence on the Fermi energy, showing a crossover as carrier density increases.

In this work, using the Boltzmann magnetotransport theory, we evaluate the LMC of WSMs along the magnetic field direction and study its dependence on the magnetic field, Fermi energy, and temperature. Contrary to conventional theories, we not only consider a pair of Weyl nodes but also a trivial band near them, demonstrating that the trivial band significantly affects the LMC. In the weak internode and node-trivial band scattering limit, where internode scatterings and scatterings between a Weyl node and the trivial band are negligible compared to intranode and intra-trivial band scatterings, the LMC is quadratic in the magnetic field, exhibiting $1 / \varepsilon_\trm{F}^{4}$ dependence at high densities due to internode scatterings and $1 / \varepsilon_\trm{F}^{2}$ dependence at low densities due to scatterings between a Weyl node and the trivial band, consistent with the experimental results in Ref.~\cite{Zhang2016}. For strong internode and node-trivial band scatterings comparable to intranode and intra-trivial band scatterings, the magnetic field-driven anisotropy arising from the phase-space volume element and the orbital magnetic moment is no longer negligible, thus the LMC substantially deviates from that in the weak internode and node-trivial band scattering limit. We also evaluate the temperature dependence of the LMC assuming strong inelastic scatterings, and find its asymptotic behaviors at low and high temperatures, respectively. The LMC exhibits $1 - \alpha T^2$ dependence for some $\alpha > 0$ in the low temperature regime, $1 / T^2$ dependence due to scatterings between a Weyl node and the trivial band in the intermediate regime, and $1 / T^4$ dependence due to internode scatterings at high temperatures.

\section{Review on Boltzmann magnetotransport theory}
\label{sec:Boltzmann_review}
In a stationary, homogeneous system with nonvanishing Berry curvature $\bm{\Omega_k}$, the Boltzmann equation states that the nonequilibrium distribution function $f_{\bm{k}}$ satisfies
\begin{equation}
	\label{eq:boltzmann_eqn}
	\frac{df_{\bm{k}}}{dt}= \dot{\bm{k}} \cdot \frac{\partial f_{\bm{k}}}{\partial \bm{k}} = \int \frac{d^d k^\prime}{(2\pi)^d} D_{\bm{k^\prime}} W_{\bm{kk^\prime}} (f_{\bm{k^\prime}} - f_{\bm{k}}),
\end{equation}
where $d$ is the dimension of the system, $D_{\bm{k}} \equiv 1 + \frac{e}{\hbar c} (\bm{\Omega_k} \cdot \bm{B})$ is the phase-space volume element \cite{Xiao2005, Xiao2010}, and $W_{\bm{kk^\prime}}$ is the transition rate from $\bm{k}$ to $\bm{k^\prime}$. If not specified otherwise, we assume nonmagnetic elastic impurity scatterings so that $W_{\bm{kk^\prime}} = \frac{2\pi}{\hbar} n_\textrm{imp} |V_{\bm{kk^\prime}}|^2 \delta(\tilde{\varepsilon}_{\bm{k}} - \tilde{\varepsilon}_{\bm{k^\prime}})$ by the Fermi golden rule, where $n_\textrm{imp}$ is the impurity density and $V_{\bm{kk^\prime}}$ is the matrix element of the impurity potential. By the magnetic field, the dispersion of electrons is modified into $\tilde{\varepsilon}_{\bm{k}} \equiv \varepsilon_{\bm{k}} - \bm{m_k} \cdot \bm{B}$, where $\varepsilon_{\bm{k}}$ and $\bm{m_k}$ are the dispersion without a magnetic field and the orbital magnetic moment, respectively.

The equations of motion for Bloch electrons read \cite{Sundaram1999}
\begin{equation}
	\label{eq:semi_eom}
	\dot{\bm{r}} = \bm{v_k} - \dot{\bm{k}}\times \bm{\Omega_k}, \qquad \hbar \dot{\bm{k}} = (-e)\left(\bm{E} + \frac{\dot{\bm{r}}}{c}  \times \bm{B}\right),
\end{equation}
where $\bm{v_k} \equiv \frac{1}{\hbar} \frac{\partial}{\partial \bm{k}} \tilde{\varepsilon}_{\bm{k}}$ is the group velocity. Thus, introducing the mean-free-path vector $\bm{L_k}$ as $f_{\bm{k}} \approx f_{\bm{k}}^\trm{eq} - e \bm{E} \cdot \bm{L_k} S (\tilde{\varepsilon}_{\bm{k}})$ to the linear order in $\bm{E}$, where $f_{\bm{k}}^\trm{eq}$ is the equilibrium Fermi-Dirac distribution and $S (\tilde{\varepsilon}_{\bm{k}}) \equiv -\partial f_{\bm{k}}^\trm{eq} / \partial \tilde{\varepsilon}_{\bm{k}}$, 
Eq.~\eqref{eq:semi_eom} transforms Eq.~\eqref{eq:boltzmann_eqn} into \cite{Knoll2020,Sharma2020, Suh2023}:
\begin{align}
	\label{eq:mfp_eqn}
	& \bm{v_k}^\trm{mod} + \frac{e}{\hbar c} \left[ (\bm{v_k}^\trm{mod} \times \bm{B}) \cdot \frac{\partial}{\partial \bm{k}} \right] \bm{L_k} \nonumber \\ & \quad = \int \frac{d^d k^\prime}{(2\pi)^d} D_{\bm{k^\prime}} W_{\bm{kk^\prime}} \left( \bm{L_k} - \bm{L_{k^\prime}} \right).
\end{align}
Here, $\bm{v_k}^\textrm{mod} \equiv D_{\bm{k}}^{-1} [\bm{v_k} + \frac{e}{\hbar c} (\bm{\Omega_k} \cdot \bm{v_k}) \bm{B}]$ is the modified velocity. Solving Eq.~\eqref{eq:mfp_eqn}, we obtain the current density from $\bm{J} = \mrm{g} (-e)\int \frac{d^d k}{(2\pi)^d} \dot{\bm{r}} f_{\bm{k}}$ and the corresponding conductivity from \cite{Suh2023}
\begin{equation}
	\label{eq:conductivity_eqn}
	\sigma_{ij} = \mathrm{g} e^2 \int \frac{d^d k}{(2\pi)^d} D_{\bm{k}} S (\tilde{\varepsilon}_{\bm k}) v_{\bm k}^{\textrm{mod}(i)} L_{\bm{k}}^{(j)}.
\end{equation}

\section{Evaluation and analysis of LMC at zero temperature}
For numerical calculations, we consider a model where a magnetic field $\bm{B} = B \hat{\bm{z}}$ is applied on a WSM with a pair of two isotropic Weyl nodes and an isotropic trivial band separated equally from each node. The effective Hamiltonian near a Weyl node is given by $H_{\bm{k}\chi} = \chi \hbar v \bm{\sigma} \cdot \bm{k}$, where $\bm{k}$ is the momentum measured from the node $\chi$ and $\bm{\sigma}$ is the vector of Pauli matrices. Without loss of generality, we assume that the Fermi energy lies on the upper band. Then, the eigenvalues are given by $\varepsilon_{\bm{k}} = \hbar v k$ with the eigenstates $| u_{\bm{k}+} \rangle = (\cos(\theta / 2), \sin(\theta / 2) e^{i\phi})^\trm{t}$ and $| u_{\bm{k}-} \rangle = (\sin(\theta / 2), -\cos(\theta / 2) e^{i\phi})^\trm{t}$ for the $\chi=\pm 1$ nodes, respectively, where $(k,\theta,\phi)$ is the spherical coordinate of $\bm{k}$. The overlap factor between the states in the Weyl nodes is given by $F_{\bm{k}\chi ; \bm{k^\prime}\chi^\prime} = \frac{1}{2} (1 + \chi\chi^\prime \hat{\bm{k}} \cdot \hat{\bm{k}}^\prime)$, the Berry curvature is given by $\bm{\Omega}_{\bm{k}\chi} = -\chi \bm{k} / 2k^3$, and the orbital magnetic moment is given by $\bm{m}_{\bm{k}\chi} = -\chi ev\bm{k} / 2ck^2$.

For the trivial band, we assume the Hamiltonian to be $H_{\bm{q}} = \varepsilon_{\bm{q}} \mathds{1}$ for some isotropic $\varepsilon_{\bm{q}}$, so that the Berry curvature and the orbital magnetic moment vanish. Here, $\bm{q}$ is the momentum measured from the center of the trivial band, and for each $\bm{q}$, there exist (pseudo)spin up/down states represented by $s = \pm 1$. The overlap factors are given by $F_{\bm{q}s ; \bm{q^\prime}s^\prime} = \delta_{ss^\prime}$ between the states in the trivial band and $F_{\bm{k}\chi ; \bm{q}s} = (1 + \chi s\cos\theta) / 2$ between the states in the Weyl node and the trivial band. In addition, we adopt the impurity potential $V_\trm{n}$ for intranode scatterings, $V_\trm{t}$ for intra-trivial band scatterings, $V_\trm{nn}$ for internode scatterings, and $V_\trm{nt}$ for scatterings between a Weyl node and the trivial band. Assuming that the Fermi wavevector is much smaller than the inverse screening length, and the separation between the Weyl nodes or between a Weyl node and the trivial band, we neglect the momentum dependence of the impurity potentials.

Based on this model, we rewrite Eq.~\eqref{eq:mfp_eqn} along the $z$-direction as
\begin{widetext}
\begin{subequations}
	\label{eq:mfp_eqn_model}
	\begin{equation}
		v^{(z)}_{\bm{q}} = \int \frac{d^3 q^\prime}{(2\pi)^3} W_{\bm{q}\bm{q^\prime}} \left( L_{\bm{q}s}^{(z)} - L_{\bm{q^\prime}s^\prime}^{(z)} \right) + \sum_\chi \int \frac{d^3 k}{(2\pi)^3} D_{\bm{k}\chi} W_{\bm{q} s; \bm{k}\chi} \left( L_{\bm{q}s}^{(z)} - L_{\bm{k}\chi}^{(z)} \right),
	\end{equation}
	and
	\begin{equation}
		v^{\trm{mod}(z)}_{\bm{k}\chi} = \sum_s \int \frac{d^3 q}{(2\pi)^3} W_{\bm{k}\chi ; \bm{q}s} \left( L_{\bm{k}\chi}^{(z)} - L_{\bm{q}s}^{(z)} \right) + \sum_{\chi^\prime} \int \frac{d^3 k^\prime}{(2\pi)^3} D_{\bm{k^\prime}\chi^\prime} W_{\bm{k}\chi ; \bm{k^\prime}\chi^\prime} \left( L_{\bm{k}\chi}^{(z)} - L_{\bm{k^\prime}\chi^\prime}^{(z)} \right).
	\end{equation}
\end{subequations}
\end{widetext}
Here, the transition rates are given by $W_{\bm{q}\bm{q^\prime}} = \frac{2\pi}{\hbar} n_\trm{imp} V_\trm{t}^2 \delta (\varepsilon_{\bm{q}} - \varepsilon_{\bm{q^\prime}})$ between the states in the trivial band, $W_{\bm{q}s ; \bm{k}\chi} = W_{\bm{k}\chi ; \bm{q}s} = \frac{2\pi}{\hbar} n_\trm{imp} V_\trm{nt}^2 F_{\bm{k}\chi ; \bm{q}s} \delta (\tilde{\varepsilon}_{\bm{k}\chi} - \varepsilon_{\bm{q}})$ between the states in the trivial band and a Weyl node, and $W_{\bm{k}\chi ; \bm{k^\prime}\chi^\prime} = \frac{2\pi}{\hbar} n_\trm{imp} V_{\chi\chi^\prime}^2 F_{\bm{k}\chi ; \bm{k^\prime}\chi^\prime} \delta (\tilde{\varepsilon}_{\bm{k}\chi} - \tilde{\varepsilon}_{\bm{k^\prime}\chi^\prime})$ between the states in the Weyl nodes, where $V_{\chi\chi^\prime} = V_\trm{n}$ for $\chi = \chi^\prime$ and $V_\trm{nn}$ otherwise. Note that the Lorentz force term, the second term on the left-hand-side of Eq.~\eqref{eq:mfp_eqn}, vanishes since it does not affect the transport along the magnetic field direction in isotropic systems. Solving Eq.~\eqref{eq:mfp_eqn_model}, we can evaluate the conductivity along the magnetic field direction through
\begin{equation}
	\label{eq:sigma_simple}
	\sigma_{zz} (B) = \sigma_{zz}^\trm{tri} (B) + \sigma_{zz}^\trm{n} (B),
\end{equation}
where
\begin{equation}
	\label{eq:sigma_tri}
	\sigma_{zz}^\trm{tri} (B) = \mrm{g} e^2 \sum_s \int \frac{d^3 q}{(2\pi)^3} S(\varepsilon_{\bm{q}}) v_{\bm{q}}^{(z)} L_{\bm{q}s}^{(z)},
\end{equation}
and
\begin{equation}
	\label{eq:sigma_n}
	\sigma_{zz}^\trm{n} (B) = \mrm{g} e^2 \sum_\chi \int \frac{d^3 k}{(2\pi)^3} D_{\bm{k}\chi} S(\tilde{\varepsilon}_{\bm{k}\chi}) v_{\bm{k}\chi}^{\trm{mod}(z)} L_{\bm{k}\chi}^{(z)}.
\end{equation}
Note that $S(\varepsilon) = \delta (\varepsilon - \varepsilon_\trm{F})$ at zero temperature. Details on the calculations are presented in Appendix \ref{sec:appen_cal_detail}.

\subsection{Dependence on the magnetic field}
\begin{figure}[t]
	\includegraphics[width=0.85\linewidth]{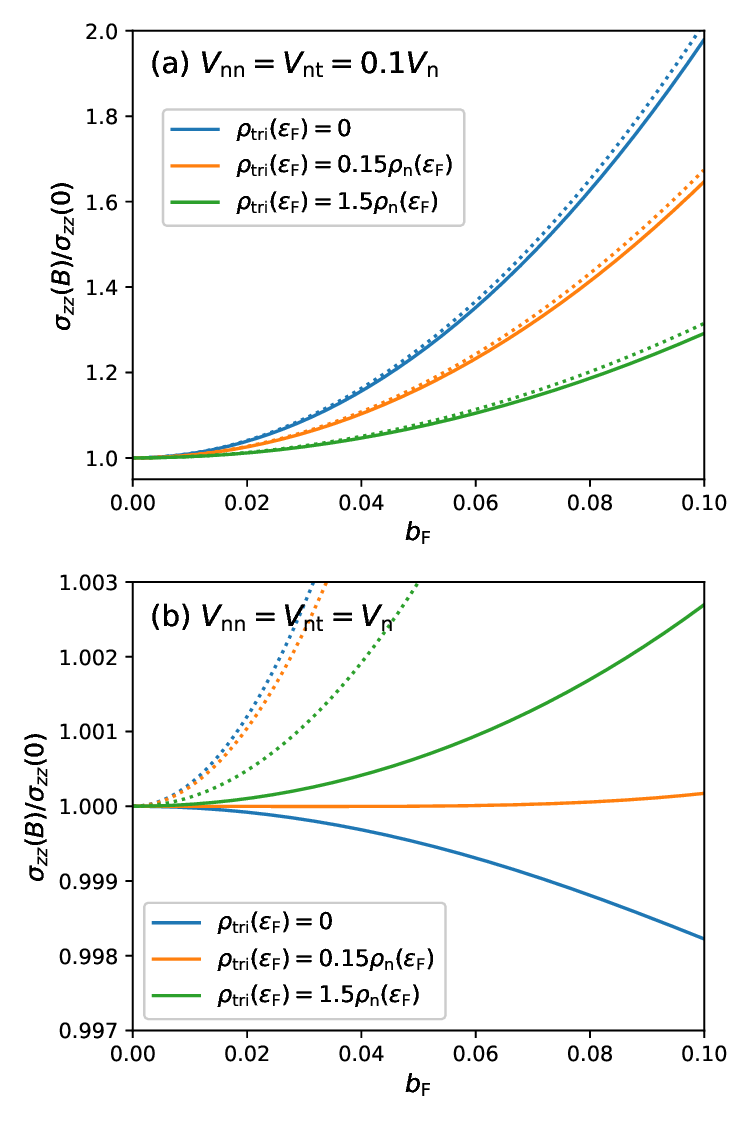}
	\caption{$\sigma_{zz} (B) / \sigma_{zz} (0)$ as a function of $b_\trm{F}$ for (a) $V_\trm{nn} = V_\trm{nt} = 0.1V_\trm{n}$ and (b) $V_\trm{nn} = V_\trm{nt} = V_\trm{n}$. The blue, orange, and green lines represent the results for $\rho_\trm{tri} (\varepsilon_\trm{F}) / \rho_\trm{n} (\varepsilon_\trm{F}) = 0, 0.15$, and $1.5$, respectively. The dashed lines with the corresponding colors represent Eq.~\eqref{eq:sigma_analytic_main} neglecting the field-driven anisotropy. Here we assume $v_\trm{t} = v$ and $V_\trm{t} = V_\trm{n}$.} 
	\label{fig:B_dep}
\end{figure}
Figure \ref{fig:B_dep} illustrates the conductivity $\sigma_{zz} (B) / \sigma_{zz} (0)$ as a function of $b_\trm{F}$, where $b_\trm{F} \equiv \frac{e}{\hbar c} |\bm{B}| |\bm{\Omega}_{\bm{k}\chi}|_{k=k_\trm{F}} = eB / 2\hbar ck_\trm{F}^2$ is the dimensionless quantity representing the coupling strength between the magnetic field and the Berry curvature \cite{Suh2023, Woo2022}. From Fig.~\ref{fig:B_dep}(a), we find that weak internode and node-trivial band scatterings yield a large positive LMC increasing quadratically with respect to the magnetic field, as seen from Eq.~\eqref{eq:LMC_approx}. As $\rho_\trm{tri} (\varepsilon_\trm{F})$ increases, where $\rho_\trm{tri} (\varepsilon_\trm{F})$ is the DOS per (pseudo)spin of the trivial band at the Fermi level, scatterings between a Weyl node and the trivial band become significant, increasing the anomalous relaxation rate and consequently lowering the LMC. On the other hand, as seen in Fig.~\ref{fig:B_dep}(b), for strong internode and node-trivial band scatterings, the negative LMC appears in the absence of the trivial band due to the phase-space volume element and the orbital magnetic moment, consistent with Refs.~\cite{Knoll2020,Sharma2020,Sharma2023}. 
As $\rho_\trm{tri} (\varepsilon_\trm{F})$ increases, however, the LMC becomes positive in contrast to the weak internode and node-trivial band scattering case.

Magnetic fields affect the LMC in two distinct ways: by inducing a charge imbalance through the chiral anomaly and by affecting intranode scatterings through the field-driven anisotropy induced by the phase-space volume element and the orbital magnetic moment. If internode and node-trivial band scatterings are sufficiently weak, the chiral charge imbalance becomes significant, making the former dominant. Otherwise, the contributions from the latter cannot be neglected. This accounts for the substantial difference between the results in Figs.~\ref{fig:B_dep}(a) and \ref{fig:B_dep}(b). For verification, we analytically obtain the conductivity neglecting the field-driven anisotropy. In the absence of the phase-space volume element and the orbital magnetic moment, Eq.~\eqref{eq:mfp_eqn} along the magnetic field direction, with no Lorentz force term, reduces to
\begin{equation}
	\label{eq:mfp_eqn_simple_main}
	v_{\bm{k}}^{(z)} + v_{\bm{k}}^{\trm{a}(z)} = \int \frac{d^d k^\prime}{(2\pi)^d} W_{\bm{kk^\prime}} \left( L_{\bm{k}}^{(z)} - L_{\bm{k^\prime}}^{(z)} \right),
\end{equation}
where $\bm{v}_{\bm{k}}^\trm{a} \equiv \frac{e}{\hbar c} (\bm{\Omega}_{\bm{k}} \cdot \bm{v_k}) \bm{B}$ is the anomalous velocity. Writing $L_{\bm{k}}^{(z)} = l_{\bm{k}}^{(z)} + l_{\bm{k}}^{\trm{a}(z)}$, we separate Eq.~\eqref{eq:mfp_eqn_simple_main} into two parts; the nonmagnetic part that does not depend on the magnetic field and the anomalous part that depends on the magnetic field. In our model, considering the isotropy of the system, we obtain $l_{\bm{q}}^{(z)} = v_{\bm{q}}^{(z)} \tau_\trm{t}^\trm{tr} (\varepsilon_{\bm{q}})$, $l_{\bm{q}s}^{\trm{a}(z)} = 0$, $l_{\bm{k}}^{(z)} = v_{\bm{k}}^{(z)} \tau_\trm{n}^\trm{tr} (\varepsilon_{\bm{k}})$, and $l_{\bm{k}\chi}^{(z)} = v_{\bm{k}\chi}^{\trm{a}(z)} \tau^\trm{a} (\varepsilon_{\bm{k}})$, where $\bm{v}_{\bm{k}\chi}^{\trm{a}} = -\chi vb_{\bm{k}} \hat{\bm{z}}$ and $b_{\bm{k}} = eB / 2\hbar c k^2$. From Eq.~\eqref{eq:mfp_eqn_model} assuming $D_{\bm{k}\chi} \approx 1$ and $\tilde{\varepsilon}_{\bm{k}\chi} \approx \varepsilon_{\bm{k}}$, the relaxation times at the Fermi energy are given by (see Appendix \ref{sec:appen_analytic_analysis} for the details)
\begin{subequations}
\label{eq:relaxation_times_main}
\begin{eqnarray}
	\frac{1}{\tau_\trm{t}^\trm{tr} (\varepsilon_\trm{F})} &=& \frac{1}{\tau_\trm{t}} + \frac{1}{\tau_\trm{nt}}, \\
	\frac{1}{\tau_\trm{n}^\trm{tr} (\varepsilon_\trm{F})} &=& \frac{1}{\tau_\trm{tn}} + \frac{1}{3\tau_\trm{n}} + \frac{2}{3\tau_\trm{nn}}, \\
	\label{eq:tau_a_main}
	\frac{1}{\tau^\trm{a} (\varepsilon_\trm{F})} &=& \frac{1}{\tau_\trm{tn}} + \frac{1}{\tau_\trm{nn}}.
\end{eqnarray}
\end{subequations}
Here, $1/\tau_\trm{n} \equiv \frac{2\pi}{\hbar} n_\trm{imp} V_\trm{n}^2 \rho_\trm{n} (\varepsilon_\trm{F})$, $1/\tau_\trm{t} \equiv \frac{2\pi}{\hbar} n_\trm{imp} V_\trm{t}^2 \rho_\trm{tri} (\varepsilon_\trm{F})$, $1/\tau_\trm{nn} \equiv \frac{2\pi}{\hbar} n_\trm{imp} V_\trm{nn}^2 \rho_\trm{n} (\varepsilon_\trm{F})$, $1/\tau_\trm{nt} \equiv \frac{2\pi}{\hbar} n_\trm{imp} V_\trm{nt}^2 \rho_\trm{n} (\varepsilon_\trm{F})$, and $1/\tau_\trm{tn} \equiv \frac{2\pi}{\hbar} n_\trm{imp} V_\trm{nt}^2 \rho_\trm{tri} (\varepsilon_\trm{F})$ characterize the scattering rates for intranode scatterings, intra-trivial band scatterings, internode scatterings, node-to-trivial band scatterings, and trivial band-to-node scatterings, respectively, and $\rho_\trm{n} (\varepsilon) = \varepsilon^2 / 2\pi^2 (\hbar v)^3$ is the DOS of a single Weyl node at energy $\varepsilon$. Finally, from Eqs.~\eqref{eq:sigma_tri} and \eqref{eq:sigma_n}, the zero temperature conductivities from the trivial band and the Weyl nodes are given by
\begin{subequations}
\label{eq:sigma_analytic_main}
\begin{equation}
	\sigma_{zz}^\trm{tri} (B) \approx 2\mrm{g}e^2 \rho_\trm{tri} (\varepsilon_\trm{F}) \left[ \frac{v_\trm{t}^2 \tau^\trm{tr}_\trm{t} (\varepsilon_\trm{F})}{3} \right] = \frac{2\sigma_0^\trm{tri}}{1 + \tau_\trm{t} / \tau_\trm{nt}}
\end{equation}
and $\sigma_{zz}^\trm{n} (B) \approx \sigma_{zz}^{\trm{n}0} (B) + \sigma_{zz}^\trm{a} (B)$, respectively, where
\begin{align}
	\label{eq:sigma_simple_node0}
	\sigma_{zz}^\trm{n0} (B) & \approx 2\mrm{g}e^2 \rho_\trm{n} (\varepsilon_\trm{F}) \left[ \frac{v^2 \tau^\trm{tr}_\trm{n} (\varepsilon_\trm{F})}{3} \right] \nonumber \\ & = 6\sigma_0^\trm{n} \left( 1 + \frac{2\tau_\trm{n}}{\tau_\trm{nn}} + \frac{3\tau_\trm{n}}{\tau_\trm{tn}} \right)^{-1}
\end{align}
is the normal conductivity that does not depend on the magnetic field, and
\begin{align}
	\label{eq:sigma_simple_anomalous}
	\sigma_{zz}^\trm{a} (B) & \approx 2 \mrm{g}e^2 \rho_\trm{n} (\varepsilon_\trm{F}) \left[ ( vb_\trm{F})^2 \tau^\trm{a} (\varepsilon_\trm{F}) \right] \nonumber \\ & = 6b_\trm{F}^2 \sigma_0^\trm{n} \left( \frac{\tau_\trm{n}}{\tau_\trm{nn}} + \frac{\tau_\trm{n}}{\tau_\trm{tn}} \right)^{-1}
\end{align}
\end{subequations}
is the anomalous conductivity that depends on the magnetic field following Eq.~\eqref{eq:LMC_approx}. Here, $\sigma_0^\trm{n} \equiv \mrm{g} e^2 \rho_\trm{n} (\varepsilon_\trm{F}) v^2 \tau_\trm{n} / 3$, $\sigma_0^\trm{tri} \equiv \mrm{g} e^2 \rho_\trm{tri} (\varepsilon_\trm{F}) v_\trm{t}^2 \tau_\trm{t} / 3$, and $v_\trm{t}$ is the Fermi velocity at the trivial band. In Fig.~\ref{fig:B_dep}, we observe that Eq.~\eqref{eq:sigma_analytic_main}, illustrated in the dotted lines, is well-fitted to the results for weak internode and node-trivial band scatterings, but shows a significant deviation from the results for strong internode and node-trivial band scatterings. Here, the deviation decreases with increasing $\rho_\trm{tri} (\varepsilon_\trm{F})$ since no field-driven anisotropy exists in the trivial band.

\subsection{Dependence on the Fermi energy}
\label{sec:Fermi_energy_dep}
\begin{figure}[t]
	\includegraphics[width=0.85\linewidth]{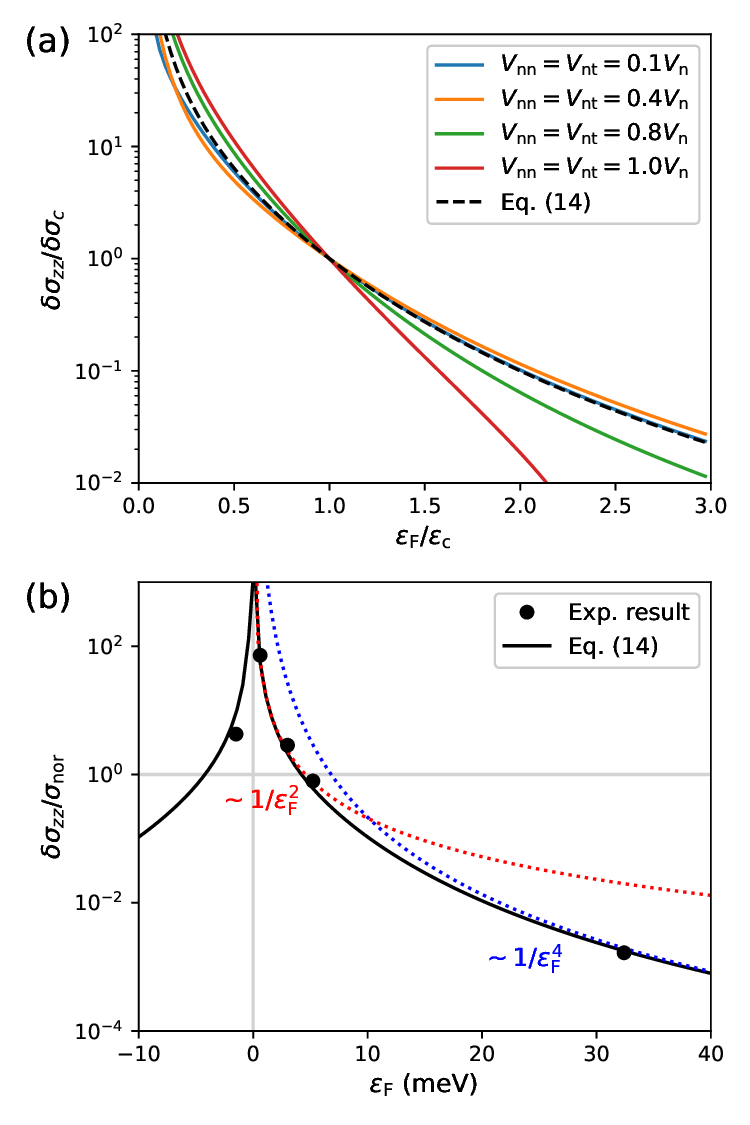}
	\caption{(a) $\delta \sigma_{zz} / \delta \sigma_\trm{c}$ as a function of $\varepsilon_\trm{F} / \varepsilon_\trm{c}$. The blue, orange, green, and red lines represent the results for $V_\trm{nn} = V_\trm{nt} = 0.1 V_\trm{n}$, $0.4 V_\trm{n}$, $0.8 V_\trm{n}$, and $1.0 V_\trm{n}$, respectively. The black dashed line represents the result obtained using Eq.~\eqref{eq:Fermi_dep_approx}. Here, we assume $b_\trm{c} \equiv eB / 2\hbar c k_\trm{c}^2 = 0.001$ with $k_\trm{c} \equiv \varepsilon_\trm{c} / \hbar v$, $v_\trm{t} = v$, and $V_\trm{t} = V_\trm{n}$. (b) Experimental data measuring the LMC of TaAs in Ref.~\cite{Zhang2016} (black dots) and its fitting curve using Eq.~\eqref{eq:Fermi_dep_approx} (black line) with $\varepsilon_\trm{c} = 10.18 \trm{ meV}$ and $\delta \sigma_\trm{c} = 0.101 \sigma_\trm{nor}$. As seen from the red and blue dotted lines, $\delta \sigma_{zz} \propto 1 / \varepsilon_\trm{F}^{2}$ at low densities and $\delta \sigma_{zz} \propto 1 / \varepsilon_\trm{F}^{4}$ at high densities. For experimental details and the definition of $\sigma_\trm{nor}$, refer to Ref.~\cite{Zhang2016}.} 
	\label{fig:fermi_energy_dep}
\end{figure}
In Fig.~\ref{fig:fermi_energy_dep}(a), we illustrate the LMC $\delta \sigma_{zz} \equiv \sigma_{zz} (B) - \sigma_{zz} (0)$ normalized by $\delta \sigma_\trm{c}$ as a function of $\varepsilon_\trm{F} / \varepsilon_\trm{c}$, where $\varepsilon_\trm{c} \equiv (V_\trm{nt} / V_\trm{nn}) \sqrt{2 \pi^2 (\hbar v)^3 \rho_\trm{tri} (0)}$ is the cross-over energy and $\delta \sigma_\trm{c} \equiv \delta \sigma_{zz} |_{\varepsilon_\trm{F} = \varepsilon_\trm{c}}$ is the LMC at $\varepsilon_\trm{F} = \varepsilon_\trm{c}$. As seen from Fig.~\ref{fig:fermi_energy_dep}(a), the LMC rapidly decreases with increasing Fermi energy due to the corresponding decrease of the Berry curvature $|\bm{\Omega}| \sim 1 / \varepsilon_\trm{F}^2$, which couples with the magnetic field. In the weak internode and node-trivial band scattering limit, $\delta \sigma_{zz} (B) \approx \sigma_{zz}^\trm{a} (B)$ given by Eq.~\eqref{eq:sigma_simple_anomalous}. From $b_\trm{F} \sim 1/\varepsilon_\trm{F}^2$, $\sigma_0^\trm{n} \sim \varepsilon_\trm{F}^0$, $\tau_\trm{n} / \tau_\trm{nn} = (V_\trm{nn} / V_\trm{n})^2 \sim \varepsilon_\trm{F}^0$, and $\tau_\trm{n} / \tau_\trm{tn} = (V_\trm{nt} / V_\trm{n})^2 [\rho_\trm{tri} (0) / \rho_\trm{n} (\varepsilon_\trm{F})] \sim 1 / \varepsilon_\trm{F}^2$, we obtain
\begin{equation}
	\label{eq:Fermi_dep_approx}
	\delta \sigma_{zz} (B) \approx \frac{2\delta \sigma_\trm{c} (B)}{(\varepsilon_\trm{F} / \varepsilon_\trm{c})^2 + (\varepsilon_\trm{F} / \varepsilon_\trm{c})^4},
\end{equation}
which is quadratic in the magnetic field from $\delta \sigma_\trm{c} (B) \propto B^2$. Here, note that near the Weyl point of our interest, we regard $\rho_\trm{tri} (\varepsilon_\trm{F}) \approx \rho_\trm{tri} (0)$ as a constant. We emphasize that at low densities where scatterings between a Weyl node and the trivial band are dominant, $\delta \sigma_{zz} \propto 1 / \varepsilon_\trm{F}^{2}$, and at high densities where internode scatterings are dominant, $\delta \sigma_{zz} \propto 1 / \varepsilon_\trm{F}^{4}$.

On the other hand, Fig.~\ref{fig:fermi_energy_dep}(b) shows the experimental data for the LMC of TaAs in Ref.~\cite{Zhang2016}. Conventional theories have claimed $\delta \sigma_{zz} \propto 1 / \varepsilon_\trm{F}^{2}$ assuming the constant anomalous relaxation time \cite{Zhang2016,Li2016}, or $\delta \sigma_{zz} \propto 1 / \varepsilon_\trm{F}^{4}$ assuming ideal WSMs with no trivial bands \cite{Parameswaran2014, Aji2012}. However, as seen in Fig.~\ref{fig:fermi_energy_dep}(b), the experimental result cannot be explained by either of the two. Indeed, since the anomalous relaxation time is much larger than the quasi-particle lifetime \cite{Zhang2016}, the sample used in the experiment belongs to the weak internode and node-trivial band scattering regime, so that the LMC would follow Eq.~\eqref{eq:Fermi_dep_approx}. From Fig.~\ref{fig:fermi_energy_dep}(b), we find that the experimental result is well-fitted to Eq.~\eqref{eq:Fermi_dep_approx}.

\section{Temperature dependence of LMC}
It is widely known that the LMC induced by the chiral anomaly rapidly decreases as the temperature increases \cite{Huang2015,Zhang2016,Li2016}. At finite temperature, inelastic scatterings such as electron-electron or electron-phonon scatterings are involved. However, exactly incorporating inelastic scatterings into the Boltzmann transport equation is quite challenging. Thus, for simplicity, we assume strong inelastic scatterings, as well as weak internode and node-trivial band scatterings, so that each Weyl node reaches local thermal equilibrium \cite{Spivak2016} with the local chemical potential $\mu_\chi = \mu - e\bm{E} \cdot \bm{l}_\chi^\trm{a}$, where $\bm{l}_\chi^\trm{a} = \langle \bm{v}_{\bm{k}\chi}^\trm{a} \rangle / \langle 1 / \tau^{\trm{a}} (\varepsilon_{\bm{k}}) \rangle$. Here, for an arbitrary $A_{\bm{k}}$, $\langle A_{\bm{k}} \rangle$ is defined by
\begin{equation}
	\label{eq:local_l_a}
	\langle A_{\bm{k}} \rangle \equiv \frac{1}{\rho_\trm{n} (\varepsilon_\trm{F})} \int d\varepsilon_{\bm{k}} \rho_\trm{n} (\varepsilon_{\bm{k}}) S(\varepsilon_{\bm{k}})A_{\bm{k}}.
\end{equation}
The corresponding LMC is given by
\begin{align}
	\label{eq:sigma_T}
	\delta \sigma_{zz} & = \mrm{g} e^2 \rho_\trm{n} (\varepsilon_\trm{F}) \sum_\chi \frac{\langle \bm{v}_{\bm{k}\chi}^\trm{a} \rangle^2}{\langle 1 / \tau^{\trm{a}} (\varepsilon_{\bm{k}}) \rangle} \\ & = \frac{\mrm{g} e^2}{4\pi^2 \hbar c} \frac{v}{c} \frac{(eB)^2 v^2}{\varepsilon_\trm{F}^2} \frac{1}{\langle 1/\tau^\trm{a} (\varepsilon_{\bm{k}}) \rangle}, \nonumber
\end{align}
which has the same form as in Eq.~\eqref{eq:LMC_approx}.
For detailed derivations, see Appendix \ref{sec:appen_finite_T}.

\begin{figure}[t]
	\includegraphics[width=0.9\linewidth]{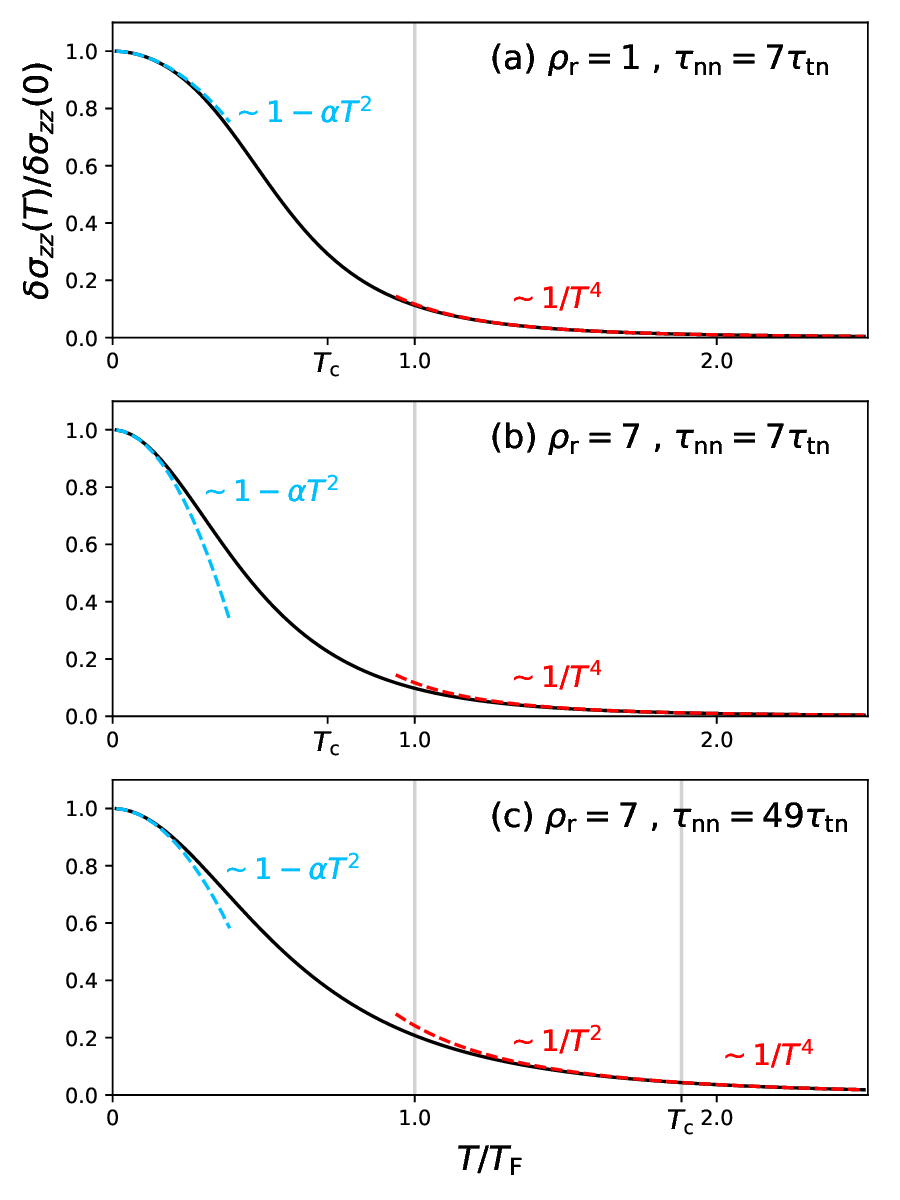}
	\caption{$\delta \sigma_{zz} (T)$ normalized by $\delta \sigma_{zz} (0)$ as a function of $T / T_\trm{F}$ for $(\rho_\trm{r}, \tau_\trm{nn} / \tau_\trm{tn})$ to be (a) $(1, 7)$, (b) $(7, 7)$, and (c) $(7,49)$, where $\rho_\trm{r} \equiv \rho_\trm{tri} (0) / \rho_\trm{n} (\varepsilon_\trm{F})$. The red and blue dashed lines represent Eqs.~\eqref{eq:high_T_asymp} and \eqref{eq:low_T_asymp}, respectively. The intermediate regime $T_\trm{F} < T < T_\trm{c}$ with $\delta\sigma_{zz} \sim 1/T^2$ only appears in (c), where the trivial band effect is significant.} 
	\label{fig:temp_dep}
\end{figure}

Figure \ref{fig:temp_dep} illustrates the LMC normalized by its value at zero temperature as a function of the normalized temperature $T / T_\trm{F}$, where $T_\trm{F}$ is the Fermi temperature. As mentioned above, the LMC rapidly decreases with temperature. To analyze further, we study the asymptotic behaviors of the LMC with the aid of Ref.~\cite{Park2017} (see Appendix \ref{sec:appen_finite_T_cal} for the details). At $T \gg T_\trm{F}$, the LMC follows
\begin{equation}
	\label{eq:high_T_asymp}
	\frac{\delta \sigma_{zz} (T)}{\delta \sigma_{zz} (0)} \approx \frac{21}{5} \frac{(\tau_\trm{tn}/\tau_\trm{nn})(1 + \tau_\trm{tn}/\tau_\trm{nn})}{(T/T_\trm{c})^2 + (T/T_\trm{c})^4},
\end{equation}
where $T_\trm{c} = \sqrt{\frac{5}{7\pi^2} (\tau_\trm{nn} / \tau_\trm{tn})} T_\trm{F}$ is the cross-over temperature. The first and second terms in the denominator on the right-hand-side of Eq.~\eqref{eq:high_T_asymp} originate from
node-trivial band scatterings
and internode scatterings, respectively. From Eq.~\eqref{eq:high_T_asymp}, $\delta\sigma_{zz} \sim 1 / T^{4}$ in the high temperature regime $T > \max (T_\trm{F}, T_\trm{c})$. If $T_\trm{c} > T_\trm{F}$, the intermediate temperature regime $T_\trm{F} < T < T_\trm{c}$ with $\delta \sigma_{zz} \sim 1 / T^{2}$ appears [Fig.~\ref{fig:temp_dep}(c)]. As observed in Figs.~\ref{fig:temp_dep}(a) and \ref{fig:temp_dep}(b) with the same value of $\tau_\trm{nn} / \tau_\trm{tn}$, the high and intermediate temperature behaviors are completely determined by $\tau_\trm{nn} / \tau_\trm{tn}$. The decrease of the LMC with temperature stems from the excited electrons at high energy where the internode and node-trivial band scattering rates are significant. Therefore, increasing $\tau_\trm{nn} / \tau_\trm{tn}$ results in a large cross-over energy $\varepsilon_\trm{c} \propto \sqrt{\tau_\trm{nn} / \tau_\trm{tn}}$, slowing down the increase of the anomalous relaxation rate with energy and consequently decelerating the decrease of the LMC with temperature. On the other hand, at $T \ll T_\trm{F}$, the LMC follows
\begin{equation}
	\label{eq:low_T_asymp}
	\frac{\delta \sigma_{zz} (T)}{\delta \sigma_{zz} (0)} \approx 1 - \alpha \left( \frac{T}{T_\trm{F}} \right)^2,
\end{equation}
where
\begin{equation}
	\alpha = \frac{\pi^2 \left[ 2(1 + 3\rho_\trm{r}) - (\tau_\trm{nn} / \tau_\trm{tn}) (1-\rho_\trm{r}) \right]}{3(1 + \rho_\trm{r})(1 + \tau_\trm{nn} / \tau_\trm{tn})}.
\end{equation}
Here, $\rho_\trm{r} \equiv \rho_\trm{tri} (0) / \rho_\trm{n} (\varepsilon_\trm{F})$. Contrary to $T > T_\trm{F}$, $\tau_\trm{nn} / \tau_\trm{tn}$ alone does not completely determine the low temperature behaviors. Given $\tau_\trm{nn} / \tau_\trm{tn}$, the LMC decreases more rapidly as $\rho_\trm{r}$ increases in the low temperature regime [Figs.~\ref{fig:temp_dep}(a) and \ref{fig:temp_dep}(b)]. At low temperatures $T \ll T_\trm{F}$, the trivial band DOS characterized by $\rho_\trm{r}$ affects the chemical potential $\mu \approx \varepsilon_\trm{F} [1 - \frac{\pi^2}{3(1 + \rho_\trm{r})} (\frac{T}{T_\trm{F}})^2]$ (see Appendix \ref{sec:appen_finite_T_cal} for its derivation) and corresponding $S(\varepsilon_{\bm{k}})$, accelerating the decrease of the LMC given by Eqs.~\eqref{eq:local_l_a} and \eqref{eq:sigma_T}. This effect barely affects the LMC at high temperatures where $\mu \approx 0$.

\section{Discussion}
This work focuses on the WSMs without tilt. However, it would be possible to analyze the tilted WSMs in a similar manner by solving Eq.~\eqref{eq:mfp_eqn} for the Hamiltonian $H_{\bm{k}\chi} = \chi \hbar v \bm{k} \cdot (\bm{\sigma} + \bm{\beta})$, where $\bm{\beta}$ is the tilt vector normalized by $v$. Here, we briefly discuss the LMC of tilted type-I WSMs with $|\bm{\beta}| < 1$ in the weak internode and node-trivial band scattering limit. Using the relaxation time approximation, Ref.~\cite{Shao2021} found that the LMC induced by the chiral anomaly can be decomposed into $\delta \sigma_{zz} (B) = \sigma_1 (B) + \sigma_2 (B)$, where
\begin{equation}
	\label{eq:sigma_1}
	\sigma_1 (B) = - \frac{\mrm{g} e^2 v}{2\pi^2 \hbar} \frac{eB}{\hbar c} \beta^{(z)} \tau^\trm{tr}_\trm{n}
\end{equation} 
with the intranode relaxation time $\tau^\trm{tr}_\trm{n}$, and
\begin{equation}
	\label{eq:sigma_2}
	\sigma_2 (B) = (1 - |\bm{\beta}|^2)^2 \frac{\mrm{g} e^2}{4\pi^2 \hbar c} \frac{v}{c} \frac{(eB)^2 v^2}{\varepsilon_\mrm{F}^2} \tau^\trm{a}
\end{equation}
with the anomalous relaxation time $\tau^\trm{a}$.
Since this type of tilt does not change the power-law dependence of the DOS and the corresponding power-law dependence of the relaxation times on the Fermi energy, $\tau_\trm{n}^\trm{tr}$ and $\tau^\trm{a}$ exhibit the same power-law dependence on the Fermi energy as that of WSMs without tilt. 
From Eq.~\eqref{eq:relaxation_times_main}, for weak internode and node-trivial band scatterings with $\tau_\trm{n} \ll \tau_\trm{nn}$ and $\tau_\trm{n} \ll \tau_\trm{tn}$, $1 / \tau^\trm{tr}_\trm{n} \sim \varepsilon_\trm{F}^{2}$, whereas $1 / \tau^\trm{a} \sim \varepsilon_\trm{F}^0$ at low densities due to scatterings between a Weyl node and the trivial band, while $1 / \tau^\trm{a} \sim \varepsilon_\trm{F}^{2}$ at high densities due to internode scatterings. Therefore, $\sigma_1 (B) \sim 1 / \varepsilon_\trm{F}^{2}$ while $\sigma_2 (B)$ follows the same power-law dependence presented in Eq.~\eqref{eq:Fermi_dep_approx}. On the other hand, for type-II WSMs with $|\bm{\beta}| > 1$, the linear Weyl Hamiltonian itself cannot capture the closed Fermi pocket. Thus, the Fermi energy dependence of the DOS, relaxation times, and LMC are determined by the higher-order terms in $\bm{k}$ in the Hamiltonian.

The Boltzmann magnetotransport theory is only valid in the semiclassical regime with a weak magnetic field $b_\trm{F} \ll 1$. To study a system with strong magnetic field or low carrier density, the quantum mechanical approach is needed, taking the Landau levels into account. We expect that contrary to conventional theories, the vertex correction should not be neglected. As seen from Eq.~\eqref{eq:tau_a_analytic} in Appendix \ref{sec:appen_analytic_analysis}, the $(1-\cos)$ factor originating from the vertex correction cancels off the contributions of intranode scatterings from the anomalous relaxation time, increasing the LMC by order $\tau_\trm{nn} / \tau_\trm{n}$. We leave the fully quantum mechanical approach for future work.

In summary, we use the Boltzmann magnetotransport theory to find how the LMC of WSMs depends on the magnetic field, Fermi energy, and temperature, incorporating the effects of the trivial band near the Weyl nodes, which have been neglected so far. The trivial band significantly affects not only the value of the LMC, but also the power-law dependence on the Fermi energy and temperature.

\acknowledgments
This work was supported by the National Research Foundation of Korea (NRF) grand funded by the Korea government (MSIT) (Grants No. 2023R1A2C1005996) and the Creative-Pioneering Researchers Program through Seoul National University (SNU).

\appendix

\onecolumngrid

\section{Details for the calculation of LMC}
\label{sec:appen_cal_detail}
In isotropic 3D WSMs, the effective Hamiltonian is given by $H_{\bm{k}\chi} = \chi \hbar v \bm{\sigma} \cdot \bm{k}$, where $\bm{k}$ is the momentum measured from a Weyl node with chirality $\chi = \pm 1$. Without loss of generality, we assume that the Fermi energy lies on the upper band. Then, the eigenvalues are given by $\varepsilon_{\bm{k}\chi} = \hbar vk$ with the eigenstates $| u_{\bm{k}+} \rangle = (\cos(\theta/2), \sin(\theta/2) e^{i\phi})^\mrm{t}$ and $| u_{\bm{k}-} \rangle = (\sin(\theta/2), -\cos(\theta/2) e^{i\phi})^\mrm{t}$  for the nodes $\chi=\pm 1$, respectively, where $(k,\theta,\phi)$ is the spherical coordinate of $\bm{k}$. The overlap factor is given by $F_{\bm{k}\chi;\bm{k^\prime}\chi^\prime} = \frac{1}{2} (1 + \chi\chi^\prime \hat{\bm{k}} \cdot \hat{\bm{k}}^\prime) = \frac{1}{2} [1 + \chi\chi^\prime (\cos\theta\cos\theta^\prime + \sin\theta\sin\theta^\prime \cos(\phi - \phi^\prime))]$, the Berry curvature is given by
\begin{subequations}
\begin{equation}
	\bm{\Omega}_{\bm{k}\chi} = -\mrm{Im} \left[ \bigg< \frac{\partial u_{\bm{k}\chi}}{\partial \bm{k}} \bigg| \times \bigg| \frac{\partial u_{\bm{k}\chi}}{\partial \bm{k}} \bigg> \right] = - \frac{\chi \hat{\bm{k}}}{2k^2},
\end{equation}
and the orbital magnetic moment is given by
\begin{equation}
	\bm{m}_{\bm{k}\chi} = \frac{e}{2\hbar c} \mrm{Im} \left[ \bigg< \frac{\partial u_{\bm{k}\chi}}{\partial \bm{k}} \bigg| \times (H_{\bm{k}\chi} - \varepsilon_{\bm{k}\chi}) \bigg| \frac{\partial u_{\bm{k}\chi}}{\partial \bm{k}} \bigg> \right] = -\frac{\chi ev\hat{\bm{k}}}{2ck}.
\end{equation}
\end{subequations}
Thus, in the presence of a magnetic field $\bm{B} = B\hat{\bm{z}}$, the phase-space volume element and the dispersion are given by $D_{\bm{k}\chi} = 1 - \chi b_\trm{F} \cos\theta / r^2$ and $\tilde{\varepsilon}_{\bm{k}\chi} = \varepsilon_\trm{F} (r + \chi b_\trm{F} \cos\theta / r)$, respectively, where $r \equiv k / k_\trm{F}$ and $b_\trm{F} \equiv eB / 2\hbar ck_\trm{F}^2$. From the dispersion, the $z$-component of the modified velocity can be written by
\begin{equation}
	v_{\bm{k}\chi}^{\trm{mod}(z)} = D_{\bm{k}\chi}^{-1} \left[ \frac{\partial \tilde{\varepsilon}_{\bm{k}\chi}}{\partial k_z} + \frac{e}{\hbar c} \left( \bm{\Omega}_{\bm{k}\chi} \cdot \frac{\partial \tilde{\varepsilon}_{\bm{k}\chi}}{\partial \bm{k}} \right) B \right] = v \left( \frac{\cos\theta - \chi b_\trm{F} \cos(2\theta) / r^2}{1 - \chi b_\trm{F} \cos\theta / r^2} - \frac{\chi b_\trm{F}}{r^2} \right).
\end{equation}
We define a new coordinate system $(R, \theta, \phi)$ with $R \equiv \tilde{\varepsilon}_{\bm{k}\chi} / \varepsilon_\trm{F} = r + \chi b_\trm{F} \cos\theta / r$ or equivalently $r_\chi (R, \theta) = R/2 + \sqrt{(R/2)^2 - \chi b_\trm{F} \cos\theta}$. The Jacobian for the coordinate transformation from $(r,\theta,\phi)$ to $(R,\theta,\phi)$ is given by
\begin{equation}
	\mathcal{J}_\chi (R,\theta) = \left| \frac{\partial r_\chi (R, \theta)}{\partial R} \right| = \frac{1}{2} \left( 1 + \frac{R/2}{\sqrt{(R/2)^2 - \chi b_\trm{F} \cos\theta}} \right).
\end{equation}

On the other hand, we assume the trivial band Hamiltonian to be $H_{\bm{q}} = \varepsilon_{\bm{q}} \mathds{1}$ for some isotropic $\varepsilon_{\bm{q}}$, where $\bm{q}$ is the momentum measured from the center of the trivial band. For each momentum $\bm{q}$, there exist (pseudo)spin up/down states represented by $s = \pm 1$. The overlap factors are given by $F_{\bm{q}s ; \bm{q^\prime} s^\prime} = \delta_{ss^\prime}$ between the states in the trivial band and $F_{\bm{k}\chi ; \bm{q}s} = \frac{1}{2} (1 + \chi s \cos\theta)$ between the states in a Weyl node and the trivial band. There are no Berry curvature or the orbital magnetic moment in the trivial band. For the scattering potential, we adopt $V_\trm{n}$ for intranode scatterings, $V_\trm{t}$ for intra-trivial band scatterings, $V_\trm{nn}$ for internode scatterings, and $V_\trm{nt}$ for node-trivial band scatterings, all of which are assumed to be momentum independent. 

Based on our model, Eq.~\eqref{eq:mfp_eqn} along the $z$-direction transforms into
\begin{subequations}
\begin{equation}
	\label{eq:mfp_eqn_model_trivial}
	v^{(z)}_{\bm{q}} = \int \frac{d^3 q^\prime}{(2\pi)^3} W_{\bm{q}\bm{q^\prime}} \left( L_{\bm{q}s}^{(z)} - L_{\bm{q^\prime}s}^{(z)} \right) + \sum_\chi \int \frac{d^3 k}{(2\pi)^3} D_{\bm{k}\chi} W_{\bm{q} s; \bm{k}\chi} \left( L_{\bm{q}s}^{(z)} - L_{\bm{k}\chi}^{(z)} \right),
\end{equation}
and
\begin{equation}
	\label{eq:mfp_eqn_model_node}
	v^{\trm{mod}(z)}_{\bm{k}\chi} = \sum_s \int \frac{d^3 q}{(2\pi)^3} W_{\bm{k}\chi ; \bm{q}s} \left( L_{\bm{k}\chi}^{(z)} - L_{\bm{q}s}^{(z)} \right) + \sum_{\chi^\prime} \int \frac{d^3 k^\prime}{(2\pi)^3} D_{\bm{k^\prime}\chi^\prime} W_{\bm{k}\chi ; \bm{k^\prime}\chi^\prime} \left( L_{\bm{k}\chi}^{(z)} - L_{\bm{k^\prime}\chi^\prime}^{(z)} \right),
\end{equation}
\end{subequations}
where the transition rates are given by $W_{\bm{q}\bm{q^\prime}} = \frac{2\pi}{\hbar} n_\trm{imp} V_\trm{t}^2 \delta (\varepsilon_{\bm{q}} - \varepsilon_{\bm{q^\prime}})$, $W_{\bm{q}s ; \bm{k}\chi} = W_{\bm{k}\chi ; \bm{q}s} = \frac{2\pi}{\hbar} n_\trm{imp} V_\trm{nt}^2 F_{\bm{k}\chi ; \bm{q}s} \delta(\tilde{\varepsilon}_{\bm{k}\chi} - \varepsilon_{\bm{q}})$, and $W_{\bm{k}\chi ; \bm{k^\prime} \chi^\prime} = \frac{2\pi}{\hbar} n_\trm{imp} V_{\chi\chi^\prime}^2 F_{\bm{k}\chi; \bm{k^\prime}\chi^\prime} \delta (\tilde{\varepsilon}_{\bm{k}\chi} - \tilde{\varepsilon}_{\bm{k^\prime}\chi^\prime})$ with $V_{\chi\chi^\prime} = V_\trm{n}$ for $\chi = \chi^\prime$ and $V_\trm{nn}$ otherwise. Here the Lorentz force term, the second term on the left-hand-side of Eq.~\eqref{eq:mfp_eqn}, vanishes since the Lorentz force does not affect the transport along the magnetic field direction in isotropic systems. In addition, since the Weyl Hamiltonian $H_{\bm{k}\chi} = \chi\hbar v \bm{k} \cdot \bm{\sigma}$ and the corresponding eigenstates are invariant under the transformation $(\bm{k},\chi) \rightarrow (-\bm{k},-\chi)$, note that $D_{\bm{k}\chi} = D_{(-\bm{k})(-\chi)}$, $W_{\bm{q}s ; \bm{k}\chi} = W_{\bm{q}s ; (-\bm{k})(-\chi)}$, and $F_{\bm{k}\chi ; \bm{k^\prime}\chi^\prime} = F_{\bm{k}\chi ; (-\bm{k^\prime})(-\chi^\prime)}$ resulting in $W_{\bm{k}\chi ; (-\bm{k^\prime})(-\chi)} = (V_\trm{nn} / V_\trm{n})^2 W_{\bm{k}\chi ; \bm{k^\prime}\chi}$. Thus, from Eq.~\eqref{eq:mfp_eqn_model_trivial}, we find $L_{\bm{q}s}^{(z)} = v_{\bm{q}}^{(z)} \tau^\trm{tr}_{\bm{q}s} + d_s (\varepsilon_{\bm{q}})$, where
\begin{equation}
	\label{eq:tau_s_tr}
	\frac{1}{\tau_{\bm{q}s}^\trm{tr}} = \int \frac{d^3 q^\prime}{(2\pi)^3} W_{\bm{qq^\prime}} \left( 1 - \hat{\bm{q}} \cdot \hat{\bm{q}}^\prime \right) + \sum_\chi \int \frac{d^3 k}{(2\pi)^3} D_{\bm{k}\chi} W_{\bm{q}s; \bm{k}\chi} = \int \frac{d^3 q^\prime}{(2\pi)^3} W_{\bm{qq^\prime}} \left( 1 - \hat{\bm{q}} \cdot \hat{\bm{q}}^\prime \right) + 2 \int \frac{d^3 k}{(2\pi)^3} D_{\bm{k}+} W_{\bm{q}s; \bm{k}+},
\end{equation}
and
\begin{equation}
	\label{eq:d_self_consistent}
	d_s (\varepsilon_{\bm{q}}) = \frac{\sum_\chi \int \frac{d^3 k}{(2\pi)^3} D_{\bm{k}\chi} W_{\bm{q}s; \bm{k}\chi} L_{\bm{k}\chi}^{(z)}}{\sum_\chi \int \frac{d^3 k}{(2\pi)^3} D_{\bm{k}\chi} W_{\bm{q}s; \bm{k}\chi}} = \frac{\int \frac{d^3 k}{(2\pi)^3} D_{\bm{k}+} W_{\bm{q}s; \bm{k}+} (L_{\bm{k}+}^{(z)} + L_{(-\bm{k})-}^{(z)})}{2\int \frac{d^3 k}{(2\pi)^3} D_{\bm{k}+} W_{\bm{q}s; \bm{k}+}}.
\end{equation}
Here, $d_s (\varepsilon_{\bm{q}})$ characterizes the asymmetry of the mean-free-path between the $\chi = +1$ and $\chi = -1$ nodes due to scatterings with the trivial band. Assuming that the trivial band is separated equally from the two Weyl nodes, we expect $d_s (\varepsilon_{\bm{q}}) = 0$ (see Appendix \ref{sec:appen_number_conservation} for the formal demonstration using the number conservation). From Eq.~\eqref{eq:mfp_eqn_model_node}, on the other hand, we obtain
\begin{align}
	\label{eq:L_self_consistent}
	L_{\bm{k}\chi}^{(z)} & = \frac{v_{\bm{k}\chi}^{\trm{mod}(z)} + \sum_{\chi^\prime} \int \frac{d^3 k^\prime}{(2\pi)^3} D_{\bm{k^\prime}\chi^\prime} W_{\bm{k}\chi; \bm{k^\prime}\chi^\prime} L_{\bm{k^\prime}\chi^\prime}^{(z)}}{\sum_s \int \frac{d^3 q}{(2\pi)^3} W_{\bm{k}\chi ; \bm{q}s} + \sum_{\chi^\prime} \int \frac{d^3 k^\prime}{(2\pi)^3} D_{\bm{k^\prime}\chi^\prime} W_{\bm{k}\chi; \bm{k^\prime}\chi^\prime}} \nonumber \\ & = \frac{v_{\bm{k}\chi}^{\trm{mod}(z)} + \int \frac{d^3 k^\prime}{(2\pi)^3} D_{\bm{k^\prime}\chi} W_{\bm{k}\chi; \bm{k^\prime}\chi} [ L_{\bm{k^\prime}\chi}^{(z)} + (V_\trm{nn} / V_\trm{n})^2 L_{(-\bm{k^\prime})(-\chi)}^{(z)}]}{\sum_s \int \frac{d^3 q}{(2\pi)^3} W_{\bm{k}\chi ; \bm{q}s} + [1 + (V_\trm{nn} / V_\trm{n})^2] \int \frac{d^3 k^\prime}{(2\pi)^3} D_{\bm{k^\prime}\chi} W_{\bm{k}\chi; \bm{k^\prime}\chi}}.
\end{align}

Assuming zero temperature, we focus on the Fermi surface by setting $R = 1$ and replacing all $\bm{k}$-dependence to the $\theta$-dependence. Then, Eq.~\eqref{eq:tau_s_tr} transforms into
\begin{equation}
	\frac{1}{\tau_s^\trm{tr} (\varepsilon_\trm{F})} = \frac{1}{\tau_\trm{t}} + \frac{1}{\tau_\trm{nt}} \int d\theta \sin\theta f(\theta) \left( \frac{1 + s\cos\theta}{2} \right),
\end{equation}
where $1 / \tau_\trm{t} \equiv \frac{2\pi}{\hbar} n_\trm{imp} V_\trm{t}^2 \rho_\trm{tri} (\varepsilon_\trm{F})$ and $1 / \tau_\trm{nt} \equiv \frac{2\pi}{\hbar} n_\trm{imp} V_\trm{nt}^2 \rho_\trm{n} (\varepsilon_\trm{F})$ characterize the intra-trivial band and node-to-trivial band scattering rates, respectively, and $\rho_\trm{n} (\varepsilon) = \varepsilon^2 / 2\pi^2 (\hbar v)^3$ and $\rho_\trm{tri} (\varepsilon)$ are the DOS of a single Weyl node and the trivial band with a single (pseudo)spin, respectively. Here, $f(\theta) \equiv f_{+} (\theta) = f_{-} (\pi - \theta)$ with $f_\chi (\theta) \equiv r_\chi^2 (1,\theta) \mathcal{J}_\chi (1, \theta) D_\chi (\theta)$, where $D_\chi (\theta) \equiv D_{\bm{k}\chi}$ on the Fermi surface. On the other hand, considering the $\theta$-dependence and $\chi$-dependence of the transition rates, we rewrite Eq.~\eqref{eq:L_self_consistent} as
\begin{equation}
	\label{eq:L_tilde_self_consistent}
	\tilde{L}_\chi^{(z)} (\theta) = \frac{\tilde{v}_{\chi}^{\trm{mod}(z)} (\theta) + \beta_{1,\chi} + \chi \beta_{2,\chi} \cos\theta}{\alpha_1 + \chi \alpha_2 \cos\theta},
\end{equation}
with
\begin{align}
	\begin{pmatrix}
		\alpha_1 \\ \alpha_2
	\end{pmatrix}
	& = \frac{\tau_\trm{n}}{\tau_\trm{tn}}
	\begin{pmatrix}
		1 \\ 0
	\end{pmatrix}
	+ \frac{1}{4} \left( 1 + \frac{\tau_\trm{n}}{\tau_\trm{nn}} \right) \int d\theta \sin\theta f_\chi (\theta)
	\begin{pmatrix}
		1 \\ \chi \cos\theta
	\end{pmatrix}
	\nonumber \\ & = \frac{\tau_\trm{n}}{\tau_\trm{tn}}
	\begin{pmatrix}
		1 \\ 0
	\end{pmatrix}
	+ \frac{1}{4} \left( 1 + \frac{\tau_\trm{n}}{\tau_\trm{nn}} \right) \int d\theta \sin\theta f (\theta)
	\begin{pmatrix}
		1 \\ \cos\theta
	\end{pmatrix},
\end{align}
and
\begin{align}
	\label{eq:beta_self_consistent_pre}
	\begin{pmatrix}
		\beta_{1,\chi} \\ \beta_{2,\chi}
	\end{pmatrix}
	& = \frac{1}{4} \sum_{\chi^\prime} \left( \frac{V_{\chi\chi^\prime}}{V_\trm{n}} \right)^2 \int d\theta^\prime \sin\theta^\prime f_{\chi^\prime} (\theta^\prime) \tilde{L}_{\chi^\prime}^{(z)} (\theta^\prime)
	\begin{pmatrix}
		1 \\ \chi^\prime \cos\theta^\prime
	\end{pmatrix}
	\nonumber \\ & = \frac{1}{4} \int d\theta \sin\theta f_{\chi} (\theta) \left[ \tilde{L}_{\chi}^{(z)} (\theta) + \frac{\tau_\trm{n}}{\tau_\trm{nn}} \tilde{L}_{-\chi}^{(z)} (\pi - \theta) \right]
	\begin{pmatrix}
		1 \\ \chi \cos\theta
	\end{pmatrix},
\end{align}
where $\tilde{L}_\chi^{(z)} (\theta) \equiv L_{\bm{k}\chi}^{(z)} / (v\tau_\trm{n})$, $\tilde{v}_\chi^{\trm{mod}(z)} (\theta) \equiv v_{\bm{k}\chi}^{\trm{mod}(z)} / v$, and $1/\tau_\trm{n} \equiv \frac{2\pi}{\hbar} n_\trm{imp} V_\trm{n}^2 \rho_\trm{n} (\varepsilon_\trm{F})$,  $1/\tau_\trm{tn} \equiv \frac{2\pi}{\hbar} n_\trm{imp} V_\trm{nt}^2 \rho_\trm{tri} (\varepsilon_\trm{F})$, and $1/\tau_\trm{nn} \equiv \frac{2\pi}{\hbar} n_\trm{imp} V_\trm{nn}^2 \rho_\trm{n} (\varepsilon_\trm{F})$ characterize the intranode, trivial band-to-node, and internode scattering rates, respectively. In general, we can write $\beta_{1,\chi} = \chi\beta_1 + \beta_1^\prime$ and $\beta_{2,\chi} = \chi\beta_2 + \beta_2^\prime$. Using the number conservation, we find $\beta_1^\prime = \beta_2^\prime = 0$ (see Appendix \ref{sec:appen_number_conservation} for details), so that Eq.~\eqref{eq:L_tilde_self_consistent} satisfies $L_{(-\bm{k})(-\chi)}^{(z)} = -L_{\bm{k}\chi}^{(z)}$ as expected from the symmetry $H_{(-\bm{k})(-\chi)} = H_{\bm{k}\chi}$ and equal separation of the trivial band from the two Weyl nodes. Thus, $\beta_1$ and $\beta_2$ can be simplified into
\begin{align}
	\label{eq:beta_self_consistent}
	\begin{pmatrix}
		\beta_{1} \\ \beta_{2}
	\end{pmatrix}
	& = \frac{1}{4} \left( 1 - \frac{\tau_\trm{n}}{\tau_\trm{nn}} \right) \int d\theta \sin\theta f_\chi (\theta) \tilde{L}_\chi^{(z)} (\theta)
	\begin{pmatrix}
		1 \\ \chi \cos\theta
	\end{pmatrix}
	\nonumber \\ & = \frac{1}{4} \left( 1 - \frac{\tau_\trm{n}}{\tau_\trm{nn}} \right) \int d\theta \sin\theta f (\theta) \tilde{L}_+^{(z)} (\theta)
	\begin{pmatrix}
		1 \\ \cos\theta
	\end{pmatrix}.
\end{align}
Substituting Eq.~\eqref{eq:L_tilde_self_consistent} into Eq.~\eqref{eq:beta_self_consistent}, we obtain the following:
\begin{equation}
	\label{eq:beta_matrix_eqn}
	\begin{pmatrix}
		\beta_{1} \\ \beta_{2}
	\end{pmatrix}
	=
	\begin{pmatrix}
		\gamma_1 \\ \gamma_2
	\end{pmatrix}
	+ \mathds{M}
	\begin{pmatrix}
		\beta_{1} \\ \beta_{2}
	\end{pmatrix}
	=
	(\mathds{1} - \mathds{M})^{-1} 
	\begin{pmatrix}
		\gamma_1 \\ \gamma_2
	\end{pmatrix},
\end{equation}
where
\begin{equation}
	\begin{pmatrix}
		\gamma_1 \\ \gamma_2
	\end{pmatrix}
	= \frac{1}{4} \left( 1 - \frac{\tau_\trm{n}}{\tau_\trm{nn}} \right) \int d\theta \frac{\sin\theta f(\theta) \tilde{v}_{+}^{\trm{mod}(z)} (\theta)}{\alpha_1 + \alpha_2 \cos\theta}
	\begin{pmatrix}
		1 \\ \cos\theta
	\end{pmatrix}
\end{equation}
and
\begin{equation}
	\mathds{M} = \frac{1}{4} \left( 1 - \frac{\tau_\trm{n}}{\tau_\trm{nn}} \right) \int d\theta \frac{\sin\theta f(\theta)}{\alpha_1 + \alpha_2 \cos\theta}
	\begin{pmatrix}
		1 & \cos\theta \\ \cos\theta & \cos^2 \theta
	\end{pmatrix}.
\end{equation}

Solving Eq.~\eqref{eq:beta_matrix_eqn}, we obtain the LMC along the magnetic field direction using Eq.~\eqref{eq:conductivity_eqn}. The conductivity $\sigma_{zz} (B) = \sigma_{zz}^\trm{tri} (B) + \sigma_{zz}^\trm{n} (B)$ is given by the sum of conductivities from the trivial band and the Weyl nodes, each of which is given by
\begin{equation}
	\label{eq:sigma_tri_numerical}
	\sigma_{zz}^\trm{tri} (B) = \mrm{g} e^2 \sum_s \int \frac{d^3 q}{(2\pi)^3} S(\varepsilon_{\bm{q}}) v_{\bm{q}}^{(z)} L_{\bm{q}s}^{(z)} = \sigma_0^\trm{tri} \sum_s \frac{\tau_{s}^\trm{tr} (\varepsilon_\trm{F})}{\tau_\trm{t}} ,
\end{equation}
and
\begin{equation}
	\label{eq:sigma_n_numerical}
	\sigma_{zz}^\trm{n} (B) = \mrm{g} e^2 \sum_\chi \int \frac{d^3 k}{(2\pi)^3} D_{\bm{k}\chi} S(\tilde{\varepsilon}_{\bm{k}\chi}) v_{\bm{k}\chi}^{\trm{mod}(z)} L_{\bm{k}\chi}^{(z)} = 3 \sigma_0^\trm{n} \int d\theta \sin\theta f (\theta) \tilde{v}_{+}^{\trm{mod} (z)} (\theta) \tilde{L}_{+}^{(z)} (\theta),
\end{equation}
where $v_\trm{t}$ is the Fermi velocity at the trivial band, and $\sigma_0^\trm{tri} \equiv \mrm{g} e^2 \rho_\trm{tri} (\varepsilon_\trm{F}) (v_\trm{t}^2 \tau_\trm{t} / 3)$ and $\sigma_0^\trm{n} \equiv \mrm{g}e^2 \rho_\trm{n} (\varepsilon_\trm{F}) (v^2 \tau_\trm{n} / 3)$ are the characteristic conductivities for the trivial band and a Weyl node, respectively. Note that since $D_{\bm{k}\chi} = D_{(-\bm{k})(-\chi)}$, $v_{\bm{k}\chi}^{\trm{mod}(z)} = - v_{(-\bm{k})(-\chi)}^{\trm{mod}(z)}$, and $L_{\bm{k}\chi}^{(z)} = - L_{(-\bm{k})(-\chi)}^{(z)}$, both Weyl nodes give the same contributions to the conductivities.

Neglecting the field-driven anisotropy, $r_\chi (R, \theta) = R$, $\mathcal{J}_\chi (R, \theta) = 1$, and $D_\chi (\theta) = 1$, then Eq.~\eqref{eq:sigma_tri_numerical} reduces to
\begin{equation}
	\sigma_{zz}^\trm{tri} (B) = \frac{2\sigma_0^\trm{tri}}{1 + \tau_\trm{t} / \tau_\trm{nt}}.
\end{equation}
In addition, $\tilde{v}_{\chi}^{\trm{mod}(z)} (\theta) = \cos\theta - \chi b_\trm{F}$, $(\alpha_1, \alpha_2)^\mrm{t} = ( \frac{1}{2} + \frac{\tau_\trm{n}}{2\tau_\trm{nn}} + \frac{\tau_\trm{n}}{\tau_\trm{tn}}, 0 )^\mrm{t}$, $\mathds{M} = \frac{1 - \tau_\trm{n}/\tau_\trm{nn}}{1 + \tau_\trm{n}/\tau_\trm{nn} + 2\tau_\trm{n}/\tau_\trm{tn}} \mrm{diag} (1, 1/3)$, $(\gamma_1 , \gamma_2)^\mrm{t} = \frac{1 - \tau_\trm{n} / \tau_\trm{nn}}{1 + \tau_\trm{n} / \tau_\trm{nn} + 2\tau_\trm{n} / \tau_\trm{tn}} (-b_\trm{F}, 1/3)^\mrm{t}$, $(\beta_1 , \beta_2)^\mrm{t} = (\frac{-b_\trm{F} (1 - \tau_\trm{n} / \tau_\trm{nn})}{2(\tau_\trm{n} / \tau_\trm{nn} + \tau_\trm{n} / \tau_\trm{tn})}, \frac{1 - \tau_\trm{n} / \tau_\trm{nn}}{2(1 + 2\tau_\trm{n}/\tau_\trm{nn} + 3\tau_\trm{n} / \tau_\trm{tn})})^\mrm{t}$, and
\begin{equation}
	\tilde{L}_\chi^{(z)} (\theta) = \frac{3\cos\theta}{1 + 2\tau_\trm{n} / \tau_\trm{nn} + 3\tau_\trm{n} / \tau_\trm{tn}} - \frac{\chi b_\trm{F}}{\tau_\trm{n} / \tau_\trm{nn} + \tau_\trm{n} / \tau_\trm{tn}},
\end{equation}
thus Eq.~\eqref{eq:sigma_n_numerical} reduces to
\begin{equation}
	\label{eq:sigma_analytic_node}
	\sigma_{zz}^\trm{n} (B) = 6\sigma_0^\trm{n} \left[ \frac{1}{1 + 2\tau_\trm{n} / \tau_\trm{nn} + 3 \tau_\trm{n} / \tau_\trm{tn}} + \frac{b_\trm{F}^2}{\tau_\trm{n} / \tau_\trm{nn} + \tau_\trm{n} / \tau_\trm{tn}} \right].
\end{equation}

\section{Constraint for the number conservation}
\label{sec:appen_number_conservation}
Restoring $d_s (\varepsilon_{\bm{q}})$ which is assumed to be zero in the previous section, Eq.~\eqref{eq:L_self_consistent} is replaced by
\begin{align}
	\label{eq:L_self_consistent_general}
	L_{\bm{k}\chi}^{(z)} & = \frac{v_{\bm{k}\chi}^{\trm{mod}(z)} + \sum_s d_s (\varepsilon_{\bm{q}}) \int \frac{d^3 q}{(2\pi)^3} W_{\bm{k}\chi ; \bm{q}s} + \sum_{\chi^\prime} \int \frac{d^3 k^\prime}{(2\pi)^3} D_{\bm{k^\prime}\chi^\prime} W_{\bm{k}\chi; \bm{k^\prime}\chi^\prime} L_{\bm{k^\prime}\chi^\prime}^{(z)}}{\sum_s \int \frac{d^3 q}{(2\pi)^3} W_{\bm{k}\chi ; \bm{q}s} + \sum_{\chi^\prime} \int \frac{d^3 k^\prime}{(2\pi)^3} D_{\bm{k^\prime}\chi^\prime} W_{\bm{k}\chi; \bm{k^\prime}\chi^\prime}} \nonumber \\ & = \frac{v_{\bm{k}\chi}^{\trm{mod}(z)} + \sum_s d_s (\varepsilon_{\bm{q}}) \int \frac{d^3 q}{(2\pi)^3} W_{\bm{k}\chi ; \bm{q}s} + \int \frac{d^3 k^\prime}{(2\pi)^3} D_{\bm{k^\prime}\chi} W_{\bm{k}\chi; \bm{k^\prime}\chi} [ L_{\bm{k^\prime}\chi}^{(z)} + (V_\trm{nn} / V_\trm{n})^2 L_{(-\bm{k^\prime})(-\chi)}^{(z)}]}{\sum_s \int \frac{d^3 q}{(2\pi)^3} W_{\bm{k}\chi ; \bm{q}s} + [1 + (V_\trm{nn} / V_\trm{n})^2] \int \frac{d^3 k^\prime}{(2\pi)^3} D_{\bm{k^\prime}\chi} W_{\bm{k}\chi; \bm{k^\prime}\chi}},
\end{align}
and consequently Eq.~\eqref{eq:beta_self_consistent_pre} is replaced by
\begin{equation}
	\label{eq:beta_self_consistent_general}
	\begin{pmatrix}
		\beta_{1,\chi} \\ \beta_{2,\chi}
	\end{pmatrix}
	= \frac{\tau_\trm{n}}{2\tau_\trm{tn}}
	\begin{pmatrix}
		\sum_s \tilde{d}_s \\ \sum_s s \tilde{d}_s
	\end{pmatrix}
	+ \frac{1}{4} \int d\theta \sin\theta f_{\chi} (\theta) \left[ \tilde{L}_{\chi}^{(z)} (\theta) + \frac{\tau_\trm{n}}{\tau_\trm{nn}} \tilde{L}_{-\chi}^{(z)} (\pi - \theta) \right]
	\begin{pmatrix}
		1 \\ \chi \cos\theta
	\end{pmatrix},
\end{equation}
where $\tilde{d}_s \equiv d_s / (v\tau_\trm{n})$. Substituting Eq.~\eqref{eq:L_tilde_self_consistent} with $\beta_{1,\chi} = \chi\beta_1 + \beta_1^\prime$ and $\beta_{2,\chi} = \chi\beta_2 + \beta_2^\prime$, we obtain Eq.~\eqref{eq:beta_matrix_eqn}, as well as
\begin{subequations}
	\label{eq:d_s_equation}
	\begin{equation}
		\tilde{d}_s \int d\theta \sin\theta f(\theta) (1 + s\cos\theta) = \int d\theta \sin\theta f(\theta) (1 + s\cos\theta) \frac{\beta_1^\prime + \beta_2^\prime \cos\theta}{\alpha_1 + \alpha_2 \cos\theta}
	\end{equation}
	from Eq.~\eqref{eq:d_self_consistent}, and
	\begin{equation}
		\begin{pmatrix}
			\beta_1^\prime \\ \beta_2^\prime
		\end{pmatrix}
		= \frac{\tau_\trm{n}}{2\tau_\trm{tn}}
		\begin{pmatrix}
			\sum_s \tilde{d}_s \\ \sum_s s \tilde{d}_s
		\end{pmatrix}
		+ \frac{1}{4} \left( 1 + \frac{\tau_\trm{n}}{\tau_\trm{nn}} \right) \int d\theta \sin\theta f(\theta) \frac{\beta_{1}^\prime + \beta_{2}^\prime \cos\theta}{\alpha_1 + \alpha_2 \cos\theta}
		\begin{pmatrix}
			1 \\ \cos\theta
		\end{pmatrix}
	\end{equation}
\end{subequations}
from Eq.~\eqref{eq:beta_self_consistent_general}. However, as seen from the invariance under the transformation $(\tilde{d}_s, \beta_1^\prime , \beta_2^\prime) \rightarrow (\tilde{d}_s + \lambda, \beta_1^\prime + \lambda \alpha_1, \beta_2^\prime + \lambda \alpha_2)$ for arbitrary $\lambda$, the four equations in Eq.~\eqref{eq:d_s_equation} are not linearly independent. Thus, to uniquely determine the mean-free-path vectors and the corresponding LMC, we introduce an additional constraint induced by the number conservation. Since the total carrier density remains the same in the absence or presence of an electric field $\bm{E}$, we have the following relation to the leading order in $\bm{E}$ (at zero temperature):
\begin{align}
	\label{eq:number_conservation}
	0 & = \sum_s \int \frac{d^3 q}{(2\pi)^3} \left( f_{\bm{q}s} - f^\trm{eq}_{\bm{q}s} \right) + \sum_\chi \int \frac{d^3 k}{(2\pi)^3} D_{\bm{k}\chi} \left( f_{\bm{k}\chi} - f_{\bm{k}\chi}^\trm{eq} \right) \nonumber \\ & = -e\bm{E} \cdot \left\{ \sum_s \int \frac{d^3 q}{(2\pi)^3} \bm{L}_{\bm{q}s} \delta (\varepsilon_{\bm{q}} - \varepsilon_\trm{F}) + \sum_\chi \int \frac{d^3 k}{(2\pi)^3} D_{\bm{k}\chi} \bm{L}_{\bm{k}\chi} \delta (\tilde{\varepsilon}_{\bm{k}\chi} - \varepsilon_\trm{F}) \right\},
\end{align}
where $f_{\bm{q}s} \approx f^\trm{eq}_{\bm{q}s} - e\bm{E} \cdot \bm{L}_{\bm{q}s} S(\varepsilon_{\bm{q}})$ and $f_{\bm{k}\chi} \approx f_{\bm{k}\chi}^\trm{eq} - e\bm{E} \cdot \bm{L}_{\bm{k}\chi} S(\tilde{\varepsilon}_{\bm{k}\chi})$ are the nonequilibrium distribution function at the trivial band and the Weyl nodes, respectively, and $f^\trm{eq}$ represents the equilibrium Fermi-Dirac distribution function. Adopting the electric field along the $z$-direction and inserting $L_{\bm{q}s}^{(z)} = v^{(z)}_{\bm{q}s} \tau^\trm{tr}_{\bm{q}s} + d_s (\varepsilon_{\bm{q}})$ and Eq.~\eqref{eq:L_tilde_self_consistent} into Eq.~\eqref{eq:number_conservation}, we have
\begin{equation}
	\label{eq:number_conservation_z}
	\rho_\trm{tri} (\varepsilon_\trm{F}) \sum_s \tilde{d}_s + \rho_\trm{n} (\varepsilon_\trm{F}) \int d\theta \sin\theta f(\theta) \frac{\beta_1^\prime + \beta_2^\prime \cos\theta}{\alpha_1 + \alpha_2 \cos\theta} = 0.
\end{equation}
Combining Eqs.~\eqref{eq:d_s_equation} and \eqref{eq:number_conservation_z}, we obtain the unique solution $\beta_1^\prime = \beta_2^\prime = \tilde{d}_s = 0$ as we expected in Appendix \ref{sec:appen_cal_detail}.

\section{Analytic analysis of LMC in the weak internode and node-trivial band scattering limit}
\label{sec:appen_analytic_analysis}
Neglecting the phase-space volume element and the orbital magnetic moment, the $z$-component of Eq.~\eqref{eq:mfp_eqn} is given by
\begin{equation}
	\label{eq:mfp_eqn_simple}
	v_{\bm{k}}^{(z)} + v_{\bm{k}}^{\trm{a}(z)} = \int \frac{d^d k^\prime}{(2\pi)^d} W_{\bm{kk^\prime}} \left( L_{\bm{k}}^{(z)} - L_{\bm{k^\prime}}^{(z)} \right),
\end{equation}
where $\bm{v_k}^\trm{a} \equiv \frac{e}{\hbar c}(\bm{\Omega_k} \cdot \bm{v_k}) \bm{B}$ is the anomalous velocity. Here the Lorentz force term which does not affect the transport along the magnetic field direction in isotropic systems is omitted. Writing $L_{\bm{k}}^{(z)} = l_{\bm{k}}^{(z)} + l_{\bm{k}}^{\trm{a}(z)}$, we naturally separate Eq.~\eqref{eq:mfp_eqn_simple} into two parts; the nonmagnetic part that does not depend on the magnetic field and the anomalous part that depends on the magnetic field.

Solving Eq.~\eqref{eq:mfp_eqn_simple} for our model, we obtain $l^{(z)}_{\bm{q}} = v^{(z)}_{\bm{q}} \tau_{\trm{t}}^\trm{tr} (\varepsilon_{\bm{q}})$ for the trivial band and $l^{(z)}_{\bm{k}} = v_{\bm{k}}^{(z)} \tau_{\trm{n}}^\trm{tr} (\varepsilon_{\bm{k}})$ for the Weyl nodes. Considering that the isotropic trivial band is placed symmetrically with respect to isotropic Weyl nodes, the relaxation times are given as follows with the well-known $(1-\cos)$ factor:
\begin{equation}
	\label{eq:tau_tr_t}
	\frac{1}{\tau^\trm{tr}_{\trm{t}} (\varepsilon_{\bm{q}})} = \int \frac{d^3 q^\prime}{(2\pi)^3} W_{\bm{qq^\prime}} (1 - \hat{\bm{q}} \cdot \hat{\bm{q}}^\prime) + \sum_\chi \int \frac{d^3 k}{(2\pi)^3} W_{\bm{q}s ; \bm{k}\chi} \left[ 1 - \frac{\bm{v_q} \cdot \bm{v_k}}{|\bm{v_q}|^2} \frac{\tau_{\trm{n}}^\trm{tr} (\varepsilon_{\bm{k}})}{\tau_{\trm{t}}^\trm{tr} (\varepsilon_{\bm{q}})} \right],
\end{equation}
which corresponds to $1/\tau^\trm{tr}_{\bm{q}s}$ in Eq.~\eqref{eq:tau_s_tr} with $D_{\bm{k}\chi} \approx 1$ and $\tilde{\varepsilon}_{\bm{k}\chi} \approx \varepsilon_{\bm{k}}$, and
\begin{equation}
	\label{eq:tau_tr_n}
	\frac{1}{\tau^\trm{tr}_{\trm{n}} (\varepsilon_{\bm{k}})} = \sum_s \int \frac{d^3 q}{(2\pi)^3} W_{\bm{k}\chi ; \bm{q}s} \left[ 1 - \frac{\bm{v_k} \cdot \bm{v_q}}{|\bm{v_k}|^2} \frac{\tau_{\trm{t}}^\trm{tr} (\varepsilon_{\bm{q}})}{\tau_{\trm{n}}^\trm{tr} (\varepsilon_{\bm{k}})} \right] + \sum_{\chi^\prime} \int \frac{d^3 k^\prime}{(2\pi)^3} W_{\bm{k}\chi ; \bm{k^\prime}\chi^\prime} (1 - \hat{\bm{k}} \cdot \hat{\bm{k}}^\prime).
\end{equation}
Note that $1/\tau^\trm{tr}_{\trm{t}}$ and $1 / \tau^\trm{tr}_{\trm{n}}$ are independent of $s$ and $\chi$, respectively. Similarly, the anomalous part also yields $l_{\bm{q}}^{\trm{a}(z)} = 0$ and $l_{\bm{k}\chi}^{\trm{a}(z)} = v_{\bm{k}\chi}^{\trm{a}(z)} \tau^\trm{a} (\varepsilon_{\bm{k}})$, where $\bm{v}_{\bm{k}\chi}^\trm{a} = -\chi v b_{\bm{k}} \hat{\bm{z}}$, $b_{\bm{k}} \equiv eB / 2\hbar c k^2$, and
\begin{equation}
	\label{eq:tau_a_analytic}
	\frac{1}{\tau^\trm{a} (\varepsilon_{\bm{k}})} = \sum_s \int \frac{d^3 q}{(2\pi)^3} W_{\bm{k}\chi ; \bm{q} s} + \sum_{\chi^\prime} \int \frac{d^3 k^\prime}{(2\pi)^3} W_{\bm{k}\chi ; \bm{k^\prime} \chi^\prime} (1 - \hat{\bm{v}}_{\bm{k}\chi}^\trm{a} \cdot \hat{\bm{v}}_{\bm{k^\prime}\chi^\prime}^\trm{a})
\end{equation}
independent on $\chi$. Here we use $\bm{v_q}^\trm{a} = \bm{0}$ due to the absence of the Berry curvature and the orbital magnetic moment in the trivial band. Note that all contributions from intranode scatterings to $\tau^\trm{a}$ vanishes by $(1 - \hat{\bm{v}}_{\bm{k}\chi}^\trm{a} \cdot \hat{\bm{v}}_{\bm{k^\prime}\chi^\prime}^\trm{a}) = (1 - \chi\chi^\prime)$ factor. Evaluating Eqs.~\eqref{eq:tau_tr_t} to \eqref{eq:tau_a_analytic} at the Fermi energy, we obtain
\begin{equation}
	\label{eq:transport_relaxation_times}
	\frac{1}{\tau_\trm{t}^\trm{tr} (\varepsilon_\trm{F})} = \frac{1}{\tau_\trm{t}} + \frac{1}{\tau_\trm{nt}},
\end{equation}
\begin{equation}
	\frac{1}{\tau_{\trm{n}}^\trm{tr} (\varepsilon_\trm{F})} = \frac{1}{\tau_\trm{tn}} + \frac{1}{2\tau_\trm{n}} \sum_{\chi^\prime} \left( \frac{V_{\chi\chi^\prime}}{V_\trm{n}} \right)^2 \int d\theta_{\bm{kk^\prime}} \sin\theta_{\bm{kk^\prime}} \left( \frac{1 + \chi\chi^\prime \cos\theta_{\bm{kk^\prime}}}{2} \right) \left( 1 - \cos\theta_{\bm{kk^\prime}} \right) = \frac{1}{\tau_\trm{tn}} + \frac{1}{3\tau_\trm{n}} + \frac{2}{3\tau_\trm{nn}},
\end{equation}
where $\theta_{\bm{kk^\prime}}$ is the angle between $\bm{k}$ and $\bm{k^\prime}$ satisfying $\cos\theta_{\bm{kk^\prime}} = \cos\theta\cos\theta^\prime + \sin\theta \sin\theta^\prime \cos(\phi - \phi^\prime)$, and
\begin{equation}
	\label{eq:anomalous_relaxation_time}
	\frac{1}{\tau^\trm{a} (\varepsilon_\trm{F})} = \frac{1}{\tau_\trm{tn}} + \frac{1}{2\tau_\trm{n}} \sum_{\chi^\prime} \left( \frac{V_{\chi\chi^\prime}}{V_\trm{n}} \right)^2 \left( 1 - \chi\chi^\prime \right) = \frac{1}{\tau_\trm{tn}} + \frac{1}{\tau_\trm{nn}}.
\end{equation}
Note that all of the relaxation times in Eqs.~\eqref{eq:transport_relaxation_times} to \eqref{eq:anomalous_relaxation_time} do not depend on $s$ and $\chi$. Approximating $\rho_\trm{tri} (\varepsilon_\trm{F}) \approx \rho_\trm{tri} (0)$, $1 / \tau^\trm{a} (\varepsilon_\trm{F}) \propto (\varepsilon_\trm{F}^2 + \varepsilon_\trm{c}^2)$, where $\varepsilon_\trm{c} \equiv (V_\trm{nt} / V_\trm{nn}) \sqrt{2 \pi^2 (\hbar v)^3 \rho_\trm{tri} (0)}$. The $\varepsilon_\trm{F}^2$ dependence in $1 / \tau^\trm{a} (\varepsilon_\trm{F})$ originates from $1/\tau_\trm{nn}$ characterizing internode scatterings, while the $\varepsilon_\trm{F}^0$ dependence originates from $1/\tau_\trm{tn}$ characterizing trivial band-to-node scatterings. On the other hand, assuming $\tau_\trm{n} \ll \tau_\trm{tn}$ and $\tau_\trm{n} \ll \tau_\trm{nn}$, $1 / \tau_\trm{n}^\trm{tr} (\varepsilon_\trm{F}) \propto \varepsilon_\trm{F}^{2}$.

When several bands cross the Fermi energy, the total conductivity is given by the sum of conductivities from each band. From the Einstein relation, we can obtain the zero-temperature conductivity through $\sigma_{zz} = \sigma_{zz}^\trm{tri} + \sigma_{zz}^\trm{n}$ with
\begin{subequations}
\label{eq:sigma_simple_alter}
\begin{equation}
	\sigma_{zz}^\trm{tri} = 2 \mrm{g} e^2 \rho_\trm{tri} (\varepsilon_\trm{F}) \mathcal{D}_\trm{tri} = \frac{2\sigma_0^\trm{tri}}{1 + \tau_\trm{t} / \tau_\trm{nt}},
\end{equation}
and
\begin{equation}
	\label{eq:sigma_simple_alter_node}
	\sigma_{zz}^\trm{n} = 2 \mrm{g} e^2 \rho_\trm{n} (\varepsilon_\trm{F}) (\mathcal{D} + \mathcal{D}_\trm{a}) = \frac{6\sigma_0^\trm{n}}{1 + 2\tau_\trm{n} / \tau_\trm{nn} + 3\tau_\trm{n} / \tau_\trm{tn}} + \frac{6\sigma_0^\trm{n} b_\trm{F}^2}{\tau_\trm{n} / \tau_\trm{nn} + \tau_\trm{n} / \tau_\trm{tn}},
\end{equation}
\end{subequations}
where $\mathcal{D}_\trm{tri} \equiv v_\trm{t}^2 \tau_{\trm{t}}^\trm{tr} (\varepsilon_\trm{F}) / 3$, $\mathcal{D} \equiv v^2 \tau_{\trm{n}}^\trm{tr} (\varepsilon_\trm{F}) / 3$, and $\mathcal{D}_\trm{a} \equiv (vb_\trm{F})^2 \tau^\trm{a} (\varepsilon_\trm{F})$ are the diffusion constants with the dimensionality 3 for the normal parts and 1 for the anomalous part. Note that Eq.~\eqref{eq:sigma_simple_alter} is exactly consistent with Eq.~\eqref{eq:sigma_analytic_main} in the main text. The first term on the right-hand-side of Eq.~\eqref{eq:sigma_simple_alter_node} corresponds to the normal conductivity of the Weyl nodes without magnetic field, while the second term corresponds to the LMC given by Eq.~\eqref{eq:LMC_approx}. Discussions regarding the $\varepsilon_\trm{F}$-dependence of the LMC are presented in Sec.~\ref{sec:Fermi_energy_dep} in the main text.

The analytic analysis in this section has assumed that the field-driven anisotropy induced by the phase-space volume element and the orbital magnetic moment is negligible. As seen in Appendix \ref{sec:appen_cal_detail}, the anisotropy yields a correction of order $b_\trm{F}$ in the velocity and the corresponding mean-free-path vector, thus $\delta l_{\bm{k}} \sim v \tau^\trm{tr} b_\trm{F}$. Assuming the weak-field limit $b_\trm{F} \ll 1$, our assumption is valid only if $\delta l_{\bm{k}} \ll l_{\bm{k}}^\trm{a} \sim vb_\trm{F} \tau^\trm{a}$, which corresponds to the weak internode and node-trivial band scattering limit $\tau_\trm{n} \ll \tau_\trm{tn}$ and $\tau_\trm{n} \ll \tau_\trm{nn}$.

\section{Anomalous mean-free-path at finite temperature}
\label{sec:appen_finite_T}
In the presence of inelastic scatterings, Eq.~\eqref{eq:mfp_eqn} along the magnetic field direction, the $z$-direction, is given as follows with the factor $(1 - f_{\bm{k^\prime}}^\trm{eq}) / (1 - f_{\bm{k}}^\trm{eq})$ \cite{Kim2019}:
\begin{equation}
	\label{eq:mfp_eqn_inelastic_general}
	v_{\bm{k}}^{\trm{mod}(z)} = \int \frac{d^3 k^\prime}{(2\pi)^3} D_{\bm{k^\prime}} W_{\bm{kk^\prime}} \left( L_{\bm{k}}^{(z)} - L_{\bm{k^\prime}}^{(z)} \right) \left( \frac{1 - f_{\bm{k^\prime}}^\trm{eq}}{1 - f_{\bm{k}}^\trm{eq}} \right),
\end{equation}
where $W_{\bm{kk^\prime}}$ is the total transition rate from $\bm{k}$ to $\bm{k^\prime}$ including the one for inelastic scatterings. Applying Eq.~\eqref{eq:mfp_eqn_inelastic_general} to our model, the anomalous part can be written by
\begin{equation}
	\label{eq:mfp_eqn_inelastic}
	v_{\bm{k}\chi}^{\trm{a}(z)} = \frac{l_{\bm{k}\chi}^{\trm{a}(z)}}{\tau^{\trm{a}} (\varepsilon_{\bm{k}})} + \int \frac{d^3 k^\prime}{(2\pi)^3} W_{\bm{kk^\prime}}^\trm{th} \left( l_{\bm{k}\chi}^{\trm{a}(z)} - l_{\bm{k^\prime}\chi}^{\trm{a}(z)} \right) \left( \frac{1 - f_{\bm{k^\prime}}^\trm{eq}}{1 - f_{\bm{k}}^\trm{eq}} \right),
\end{equation}
where $v_{\bm{k}\chi}^{\trm{a}(z)} = - \chi v b_{\bm{k}}$, $\tau^{\trm{a}}$ is the anomalous relaxation time at zero temperature given by Eq.~\eqref{eq:tau_a_analytic} and $W_{\bm{kk^\prime}}^\trm{th}$ is the transition rate for inelastic scatterings from $\bm{k}$ to $\bm{k^\prime}$. For elastic scatterings, we assume the weak internode and node-trivial band scattering limit, incorporating their contributions into the first term on the right-hand-side of Eq.~\eqref{eq:mfp_eqn_inelastic} and neglecting all the field-driven anisotropy. For inelastic scatterings, we only consider the intranode contributions for simplicity. From Eq.~\eqref{eq:mfp_eqn_inelastic}, we obtain
\begin{equation}
	\label{eq:l_a_form}
	l_{\bm{k}\chi}^{\trm{a}(z)} = \frac{l_{\bm{k}\chi}^{\trm{th}(z)} + v_{\bm{k}\chi}^{\trm{a}(z)} \tau_{\bm{k}}^\trm{th}}{1 + \tau_{\bm{k}}^\trm{th} / \tau^{\trm{a}} (\varepsilon_{\bm{k}})} ,
\end{equation}
where $\tau^\trm{th}$ is the quasiparticle lifetime for inelastic scatterings given by
\begin{subequations}
\begin{equation}
	\frac{1}{\tau_{\bm{k}}^\trm{th}} = \int \frac{d^3 k^\prime}{(2\pi)^3} W_{\bm{kk^\prime}}^\trm{th} \left( \frac{1 - f_{\bm{k^\prime}}^\trm{eq}}{1 - f_{\bm{k}}^\trm{eq}} \right),
\end{equation}
and $l^{\trm{th}(z)}$ is defined by
\begin{equation}
	\label{eq:tau_th_def}
	\frac{l_{\bm{k}\chi}^{\trm{th}(z)}}{\tau_{\bm{k}}^\trm{th}} \equiv \int \frac{d^3 k^\prime}{(2\pi)^3} W_{\bm{kk^\prime}}^\trm{th} \left( \frac{1 - f_{\bm{k^\prime}}^\trm{eq}}{1 - f_{\bm{k}}^\trm{eq}} \right) l_{\bm{k^\prime}\chi}^{\trm{a}(z)}.
\end{equation}
\end{subequations}
Combining Eqs.~\eqref{eq:l_a_form} and \eqref{eq:tau_th_def}, we have
\begin{equation}
	\label{eq:mfp_eqn_inelastic_alter}
	\int \frac{d^3 k^\prime}{(2\pi)^3} W_{\bm{kk^\prime}}^\trm{th} \left( \frac{1 - f_{\bm{k^\prime}}^\trm{eq}}{1 - f_{\bm{k}}^\trm{eq}} \right) \left[ l_{\bm{k}\chi}^{\trm{th}(z)} - \frac{l_{\bm{k^\prime}\chi}^{\trm{th}(z)}}{1 + \tau_{\bm{k^\prime}}^\trm{th} / \tau^{\trm{a}} (\varepsilon_{\bm{k^\prime}})} \right] = \int \frac{d^3 k^\prime}{(2\pi)^3} W_{\bm{kk^\prime}}^\trm{th} \left( \frac{1 - f_{\bm{k^\prime}}^\trm{eq}}{1 - f_{\bm{k}}^\trm{eq}} \right) \frac{v_{\bm{k^\prime}\chi}^{\trm{a}(z)} \tau_{\bm{k^\prime}}^\trm{th}}{1 + \tau_{\bm{k^\prime}}^\trm{th} / \tau^{\trm{a}} (\varepsilon_{\bm{k^\prime}})}.
\end{equation}
Solving Eq.~\eqref{eq:mfp_eqn_inelastic_alter}, we can evaluate the anomalous mean-free-path and the corresponding conductivity.

In this work, we focus on the strong inelastic scattering limit $\tau^\trm{th} \ll \tau^{\trm{a}0}$. To the 0th order in $\tau^\trm{th}$, Eq.~\eqref{eq:mfp_eqn_inelastic_alter} reduces to
\begin{equation}
	\label{eq:mfp_eqn_alter_0th}
	\int \frac{d^3 k^\prime}{(2\pi)^3} W_{\bm{kk^\prime}}^\trm{th} \left( \frac{1 - f_{\bm{k^\prime}}^\trm{eq}}{1 - f_{\bm{k}}^\trm{eq}} \right) \left( l_{\bm{k}\chi}^{\trm{th}(z)} - l_{\bm{k^\prime}\chi}^{\trm{th}(z)} \right) = 0.
\end{equation}
From Eq.~\eqref{eq:mfp_eqn_alter_0th}, we find $l_{\bm{k}\chi}^{\trm{th}(z)} \approx l_{\chi}^{\trm{th}(z)}$ independent of $\bm{k}$. Note that for an arbitrary $G_{\bm{k}}$, using $S(\varepsilon_{\bm{k}}) = \beta f_{\bm{k}}^\trm{eq} (1 - f_{\bm{k}}^\trm{eq})$ and the detailed balance $W_{\bm{kk^\prime}}^\trm{th} f_{\bm{k}}^\trm{eq} (1 - f_{\bm{k^\prime}}^\trm{eq}) = W_{\bm{k^\prime k}}^\trm{th} f_{\bm{k^\prime}}^\trm{eq} (1 - f_{\bm{k}}^\trm{eq})$, we obtain the following relation:
\begin{align}
	\label{eq:thermal_int_formula}
	& \int \frac{d^3 k}{(2\pi)^3} S(\varepsilon_{\bm{k}}) \left[ \int \frac{d^3 k^\prime}{(2\pi)^3} W_{\bm{kk^\prime}}^\trm{th} \left( \frac{1 - f_{\bm{k^\prime}}^\trm{eq}}{1 - f_{\bm{k}}^\trm{eq}} \right) G_{\bm{k^\prime}} \right] = \beta \int \frac{d^3 k}{(2\pi)^3} \int \frac{d^3 k^\prime}{(2\pi)^3} W_{\bm{kk^\prime}}^\trm{th} f_{\bm{k}}^\trm{eq} \left( 1 - f_{\bm{k^\prime}}^\trm{eq} \right) G_{\bm{k^\prime}} \nonumber \\ & = \beta \int \frac{d^3 k^\prime}{(2\pi)^3} G_{\bm{k^\prime}} \int \frac{d^3 k}{(2\pi)^3} W_{\bm{k^\prime k}}^\trm{th} f_{\bm{k^\prime}}^\trm{eq} \left( 1 - f_{\bm{k}}^\trm{eq} \right) = \int \frac{d^3 k^\prime}{(2\pi)^3} \beta f_{\bm{k^\prime}}^\trm{eq} \left( 1 - f_{\bm{k^\prime}}^\trm{eq} \right) G_{\bm{k^\prime}} \int \frac{d^3 k}{(2\pi)^3} W_{\bm{k^\prime k}}^\trm{th} \left( \frac{1 - f_{\bm{k}}^\trm{eq}}{1 - f_{\bm{k^\prime}}^\trm{eq}} \right) \nonumber \\ & = \int \frac{d^3 k^\prime}{(2\pi)^3} S(\varepsilon_{\bm{k^\prime}}) G_{\bm{k^\prime}} / \tau^\trm{th}_{\bm{k^\prime}}.
\end{align}
With the aid of Eq.~\eqref{eq:thermal_int_formula}, Eq.~\eqref{eq:mfp_eqn_inelastic_alter} transforms into
\begin{equation}
	\label{eq:l_a_intermediate}
	l_{\chi}^{\trm{th}(z)} \int d\varepsilon_{\bm{k}} \frac{\rho_\trm{n} (\varepsilon_{\bm{k}}) S(\varepsilon_{\bm{k}}) / \tau^{\trm{a}} (\varepsilon_{\bm{k}})}{1 + \tau^\trm{th}_{\bm{k}} / \tau^{\trm{a}} (\varepsilon_{\bm{k}})} = \int d\varepsilon_{\bm{k}} \frac{ \rho_\trm{n} (\varepsilon_{\bm{k}}) S(\varepsilon_{\bm{k}}) v_{\bm{k}\chi}^{\trm{a}(z)}}{1 + \tau^\trm{th}_{\bm{k}} / \tau^{\trm{a}} (\varepsilon_{\bm{k}})}.
\end{equation}
Comparing both sides of Eq.~\eqref{eq:l_a_intermediate}, we find $l^{\trm{th}(z)} \sim vb\tau^\trm{a}$. To the leading order in $\tau^\trm{th} / \tau^{\trm{a}} \ll 1$, Eq.~\eqref{eq:l_a_form} results in $l_{\chi}^{\trm{a}(z)} \approx l_{\chi}^{\trm{th}(z)} \approx \langle v_{\bm{k}\chi}^{\trm{a}(z)} \rangle / \langle 1 / \tau^{\trm{a}} (\varepsilon_{\bm{k}}) \rangle$ independent of $\bm{k}$, where $\langle A_{\bm{k}} \rangle$ for an arbitrary $A_{\bm{k}}$ is defined by
\begin{equation}
	\left< A_{\bm{k}} \right> \equiv \frac{1}{\rho_\trm{n} (\varepsilon_\trm{F})} \int d\varepsilon_{\bm{k}} \rho_\trm{n} (\varepsilon_{\bm{k}}) S(\varepsilon_{\bm{k}}) A_{\bm{k}}.
\end{equation}

Finally, we can obtain the LMC through Eq.~\eqref{eq:conductivity_eqn}. Neglecting the field-driven anisotropy, the anomalous contribution corresponding to the LMC is given by
\begin{equation}
	\label{eq:sigma_finite_T}
	\sigma_{zz}^\trm{a} (B) \approx \mrm{g} e^2 \sum_\chi \int d\varepsilon_{\bm{k}} \rho_\trm{n} (\varepsilon_{\bm{k}}) S(\varepsilon_{\bm{k}}) v_{\bm{k}\chi}^{\trm{a}(z)} l_{\bm{k}\chi}^{\trm{a}(z)} \approx \mrm{g} e^2 \rho_\trm{n} (\varepsilon_\trm{F}) \sum_\chi \frac{\langle v_{\bm{k}\chi}^{\trm{a}(z)} \rangle^2}{\langle 1 / \tau^{\trm{a}} (\varepsilon_{\bm{k}}) \rangle} = \frac{\mrm{g} e^2}{4\pi^2 \hbar c} \frac{v}{c} \frac{(eB)^2 v^2}{\varepsilon_\trm{F}^2} \frac{1}{\langle 1 / \tau^\trm{a} (\varepsilon_{\bm{k}}) \rangle}.
\end{equation}

Due to strong intranode scatterings, the normal contribution to the distribution function is negligible. Thus, focusing on the anomalous contribution, the distribution function deviates from the equilibrium Fermi-Dirac distribution by $\delta f_{\bm{k}\chi} \approx -e \bm{E} \cdot \bm{l}_{\chi}^{\trm{a}} S(\varepsilon_{\bm{k}})$ with a $\bm{k}$-independent $\bm{l}_\chi^\trm{a} \equiv l_{\chi}^{\trm{a}(z)} \hat{\bm{z}}$. From $S(\varepsilon_{\bm{k}}) = - \partial f_{\bm{k}}^\trm{eq} / \partial \varepsilon_{\bm{k}}$, we have
\begin{equation}
	f_{\bm{k}\chi} \approx f^\trm{eq} (\varepsilon_{\bm{k}} - \mu) + e\bm{E} \cdot \bm{l}_{\chi}^{\trm{a}} \frac{\partial}{\partial \varepsilon_{\bm{k}}} f^\trm{eq} (\varepsilon_{\bm{k}} - \mu) \approx f^\trm{eq} (\varepsilon_{\bm{k}} - (\mu - e\bm{E} \cdot \bm{l}_{\chi}^{\trm{a}})),
\end{equation}
where $f^\trm{eq} (\varepsilon) \equiv 1 / (e^{\beta\varepsilon} + 1)$ satisfying $f_{\bm{k}}^\trm{eq} = f^\trm{eq} (\varepsilon_{\bm{k}} - \mu)$, so that each node reaches local thermal equilibrium with the local chemical potential $\mu_\chi = \mu - e\bm{E} \cdot \bm{l}_{\chi}^{\trm{a}}$.

\section{Asymptotic form of LMC at finite temperature}
\label{sec:appen_finite_T_cal}
\subsection{Chemical potential}
Since the carrier density measured from the charge neutral point does not vary under the temperature change, we obtain
\begin{equation}
	\label{eq:carrier_density}
	n = \int_0^{\varepsilon_\trm{F}} d\varepsilon \rho(\varepsilon) = \int_0^\infty d\varepsilon \rho(\varepsilon) \left[ f^\trm{eq} (\varepsilon - \mu) - f^\trm{eq} (\varepsilon + \mu) \right],
\end{equation} 
where $\rho (\varepsilon) = 2 \rho_\trm{tri} (\varepsilon) + 2 \rho_\trm{n} (\varepsilon)$ is the total DOS of the model. Here we used $f^\trm{eq}(\varepsilon + \mu) + f^\trm{eq} (-\varepsilon - \mu) = 1$. Regarding $\rho_\trm{tri} (\varepsilon) \approx \rho_\trm{tri} (0)$ as a constant near the Weyl point energy, $\rho(\varepsilon) \approx (\varepsilon^2 + \varepsilon_0^2) / \pi^2 (\hbar v)^3$, where $\varepsilon_0 \equiv \sqrt{2 \pi^2 (\hbar v)^3 \rho_\trm{tri} (0)} = \sqrt{\rho_\trm{r}} \varepsilon_\trm{F}$, where $\rho_\trm{r} \equiv \rho_\trm{tri} (0) / \rho_\trm{n} (\varepsilon_\trm{F})$. To proceed further, we introduce the following integral \cite{Park2017}:
\begin{align}
	\label{eq:int_formula}
	\int_0^\infty dx \frac{x^{\alpha - 1}}{z^{-1} e^x + 1} & = \int_0^\infty \frac{x^{\alpha - 1} z e^{-x}}{1 + z e^{-x}} = -\int_0^\infty dx x^{\alpha - 1} \sum_{n = 1}^{\infty} (-z)^n e^{-nx} \nonumber \\ & \overset{(t = nx)}{=} \left[ \int_0^\infty dt t^{\alpha - 1} e^{-t} \right] \left[ -\sum_{n=1}^\infty \frac{(-z)^n}{n^\alpha} \right] = \Gamma(\alpha) F_\alpha (z),
\end{align}
where $\Gamma (\alpha)$ is the gamma function and $F_\alpha (z) \equiv -\sum_{n=1}^{\infty} \frac{(-z)^n}{n^\alpha}$. 
Note that $\Gamma (\alpha)=(\alpha-1)\Gamma(\alpha-1)$ and $F_1 (z)=\ln (1+z)$.
With the aid of Eq.~\eqref{eq:int_formula}, Eq.~\eqref{eq:carrier_density} transforms into
\begin{equation}
	\label{eq:mu_eqn}
	\frac{F_3 (z) - F_3 (1/z)}{(\beta\varepsilon_\trm{F})^3} + \frac{\rho_\trm{r}}{2} \left( \frac{\mu}{\varepsilon_\trm{F}} \right) = \frac{1 + 3\rho_\trm{r}}{6},
\end{equation}
where $z \equiv e^{\beta\mu}$. At low temperature, the Sommerfeld expansion reads \cite{Ashcroft1976}
\begin{equation}
	\label{eq:Sommerfeld}
	\lim_{z \rightarrow \infty} \int_0^\infty dx \frac{H(x)}{z^{-1} e^x + 1} \approx \int_0^{\beta\mu} dx H(x) + \frac{\pi^2}{6} H^\prime (\beta\mu),
\end{equation}
where $H(x)$ is a function diverging no more rapidly than a polynomial as $x \rightarrow \infty$. 
Then using \eqref{eq:int_formula} for $H(x) = x^{\alpha-1}$, Eq. (\ref{eq:Sommerfeld}) becomes
\begin{equation}
	\label{eq:F_z_infty}
	\lim_{z\rightarrow \infty} F_{\alpha}(z)\approx {(\beta\mu)^{\alpha}\over \Gamma(\alpha+1)}\left[1+{\pi^2\over 6}{\alpha(\alpha-1)\over(\beta\mu)^2}\right],
\end{equation}
whereas $F_{\alpha}(z^{-1})=z^{-1}-{z^{-2}\over 2^{\alpha}}+\cdots$ vanishes as $z\rightarrow \infty$.
Thus from Eq.~\eqref{eq:mu_eqn}, we obtain
\begin{equation}
	\label{eq:mu_low_T}
	\frac{\mu}{\varepsilon_\trm{F}} \approx 1 - \frac{\pi^2}{3(1 + \rho_\trm{r})} \left( \frac{T}{T_\trm{F}} \right)^2 \quad \text{if } T \ll T_\trm{F}.
\end{equation}
On the other hand, at high temperature, $\beta\mu\rightarrow 0$ due to the finite carrier densities, thus $z\rightarrow 1$. From $z \approx 1 + \beta\mu + (\beta\mu)^2 / 2$ for $|\beta\mu| \ll 1$,
\begin{equation}
	\label{eq:F_high_T}
	\lim_{z \rightarrow 1} F_\alpha (z) \approx \eta(\alpha) + \eta(\alpha-1) \beta\mu + \frac{\eta(\alpha-2)}{2} (\beta\mu)^2,
\end{equation}
where $\eta(\alpha) \equiv F_\alpha (1)$ is the Dirichlet eta function \cite{Arfken2012}. Substituting Eq.~\eqref{eq:F_high_T} into Eq.~\eqref{eq:mu_eqn}, we obtain
\begin{equation}
	\label{eq:mu_high_T}
	\frac{\mu}{\varepsilon_\trm{F}} \approx \frac{(1 + 3\rho_\trm{r}) (T_\trm{F} / T)^2}{\pi^2 + 3\rho_\trm{r} (T_\trm{F} / T)^2} \quad \text{if } T \gg T_\trm{F}.
\end{equation}
Here we used $\eta(2)=\pi^2/12$.

\subsection{Anomalous conductivity}
Solving Eq.~\eqref{eq:tau_a_analytic} at arbitrary energy $\varepsilon_{\bm{k}}$, we have
\begin{equation}
	\frac{1}{\tau^\trm{a} (\varepsilon_{\bm{k}})} \approx \frac{1}{\tau_\trm{nn} (\varepsilon_{\bm{k}})} + \frac{1}{\tau_\trm{tn} (\varepsilon_{\bm{k}})} = \frac{1}{\tau_\trm{nn}} \left( \frac{\varepsilon_{\bm{k}}}{\varepsilon_\trm{F}} \right)^2 + \frac{1}{\tau_\trm{tn}},
\end{equation}
where $1 / \tau_\trm{nn} (\varepsilon_{\bm{k}}) \equiv \frac{2\pi}{\hbar} n_\trm{imp} V_\trm{nn}^2 \rho_\trm{n} (\varepsilon_{\bm{k}})$ and $1 / \tau_\trm{tn} (\varepsilon_{\bm{k}}) \equiv \frac{2\pi}{\hbar} n_\trm{imp} V_\trm{nt}^2 \rho_\trm{tri} (0)$ with $\rho_\trm{tri} (\varepsilon_{\bm{k}}) \approx \rho_\trm{tri} (0)$. Therefore, $\langle 1 / \tau^\trm{a} (\varepsilon_{\bm{k}}) \rangle$ in Eq.~\eqref{eq:sigma_finite_T} is given by
\begin{equation}
	\left< \frac{1}{\tau^\trm{a} (\varepsilon_{\bm{k}})} \right> = \frac{1}{\rho_\trm{n} (\varepsilon_\trm{F})} \int d\varepsilon_{\bm{k}} \frac{\rho_\trm{n} (\varepsilon_{\bm{k}}) S(\varepsilon_{\bm{k}})}{\tau^\trm{a} (\varepsilon_{\bm{k}})} = \frac{1}{\tau_\trm{nn}} \left[ \frac{1}{\varepsilon_\trm{F}^4} \int d\varepsilon_{\bm{k}} S ( \varepsilon_{\bm{k}} ) ( \varepsilon_{\bm{k}}^4 + \varepsilon_\trm{c}^2 \varepsilon_{\bm{k}}^2 ) \right],
\end{equation}
where $\varepsilon_\trm{c} \equiv (V_\trm{nt} / V_\trm{nn}) \varepsilon_0 = \sqrt{\tau_\trm{nn} / \tau_\trm{tn}} \varepsilon_\trm{F}$ is the cross-over energy. Utilizing $\frac{\partial}{\partial \varepsilon} f^\trm{eq} (-\varepsilon - \mu) = - \frac{\partial}{\partial \varepsilon} f^\trm{eq} (\varepsilon + \mu)$, we obtain the following with the aid of Eq.~\eqref{eq:int_formula}:
\begin{align}
	\int_{-\infty}^{\infty} d\varepsilon_{\bm{k}} S ( \varepsilon_{\bm{k}} ) ( \varepsilon_{\bm{k}}^4 + \varepsilon_\trm{c}^2 \varepsilon_{\bm{k}}^2 ) & = \int_{0}^{\infty} d\varepsilon_{\bm{k}} ( \varepsilon_{\bm{k}}^4 + \varepsilon_\trm{c}^2 \varepsilon_{\bm{k}}^2 ) \left\{ -\frac{\partial}{\partial \varepsilon_{\bm{k}}} \left[ f^\trm{eq} (\varepsilon_{\bm{k}} - \mu) + f^\trm{eq} (\varepsilon_{\bm{k}} + \mu) \right] \right\} \nonumber \\ & \quad =  \int_{0}^{\infty} d\varepsilon_{\bm{k}} ( 4 \varepsilon_{\bm{k}}^3 + 2 \varepsilon_\trm{c}^2 \varepsilon_{\bm{k}} ) \left[ f^\trm{eq} (\varepsilon_{\bm{k}} - \mu) + f^\trm{eq} (\varepsilon_{\bm{k}} + \mu) \right] \nonumber \\ & = 24(k_\trm{B} T)^4 \left[ F_4(z) + F_4 (1/z) \right] + 2 (k_\trm{B} T)^2 \varepsilon_\trm{c}^2 \left[ F_2(z) + F_2 (1/z) \right],
\end{align}
resulting in
\begin{equation}
	\frac{\langle 1 / \tau^\trm{a} (\varepsilon_{\bm{k}}) \rangle}{1 / \tau^\trm{a} (\varepsilon_\trm{F})} = \frac{\tau_\trm{tn}}{\tau_\trm{tn} + \tau_\trm{nn}} \left\{ 24 \left( \frac{T}{T_\trm{F}} \right)^4 \left[  F_4(z) + F_4 (1/z) \right] + 2 \left( \frac{\tau_\trm{nn}}{\tau_\trm{tn}} \right) \left( \frac{T}{T_\trm{F}} \right)^2 \left[  F_2(z) + F_2 (1/z) \right] \right\}.
\end{equation}
At $T \ll T_\trm{F}$, using Eqs.~\eqref{eq:F_z_infty} and \eqref{eq:mu_low_T}, we obtain
\begin{equation}
	\label{eq:energy_int_low_T}
	\frac{\langle 1 / \tau^\trm{a} (\varepsilon_{\bm{k}}) \rangle}{1 / \tau^\trm{a} (\varepsilon_\trm{F})} \approx 1 + \frac{\pi^2 \left[ 2(1 + 3\rho_\trm{r}) - (\tau_\trm{nn} / \tau_\trm{tn}) (1-\rho_\trm{r}) \right]}{3(1 + \rho_\trm{r})(1 + \tau_\trm{nn} / \tau_\trm{tn})} \left( \frac{T}{T_\trm{F}} \right)^2.
\end{equation}
On the other hand, at $T \gg T_\trm{F}$, using Eq.~\eqref{eq:F_high_T} with $\eta(2) = \pi^2 / 12$ and $\eta(4) = 7\pi^4 / 720$, we have
\begin{equation}
	\label{eq:energy_int_high_T}
	\frac{\langle 1 / \tau^\trm{a} (\varepsilon_{\bm{k}}) \rangle}{1 / \tau^\trm{a} (\varepsilon_\trm{F})} \approx \frac{\tau_\trm{tn}}{\tau_\trm{nn} + \tau_\trm{tn}} \frac{7\pi^4}{15} \left( \frac{T}{T_\trm{F}} \right)^4 + \frac{\tau_\trm{nn}}{\tau_\trm{nn} + \tau_\trm{tn}} \frac{\pi^2}{3} \left( \frac{T}{T_\trm{F}} \right)^2.
\end{equation}
Finally, inserting Eqs.~\eqref{eq:energy_int_low_T} and \eqref{eq:energy_int_high_T} into Eq.~\eqref{eq:sigma_finite_T}, the asymptotic forms of the anomalous conductivity at low and high temperatures, respectively, are given by
\begin{equation}
	\sigma_{zz}^\trm{a} (T) \approx \sigma_{zz}^\trm{a} (0) \left\{ 1 - \frac{\pi^2 \left[ 2(1 + 3\rho_\trm{r}) - (\tau_\trm{nn} / \tau_\trm{tn}) (1-\rho_\trm{r}) \right]}{3(1 + \rho_\trm{r})(1 + \tau_\trm{nn} / \tau_\trm{tn})} \left( \frac{T}{T_\trm{F}} \right)^2 \right\}
\end{equation}
for $T \ll T_\trm{F}$, and
\begin{equation}
	\label{eq:high_T_asymp_appen}
	\sigma_{zz}^\trm{a} (T) \approx \sigma_{zz}^\trm{a} (0) \left[ \frac{\tau_\trm{tn}}{\tau_\trm{nn} + \tau_\trm{tn}} \frac{7\pi^4}{15} \left( \frac{T}{T_\trm{F}} \right)^4 + \frac{\tau_\trm{nn}}{\tau_\trm{nn} + \tau_\trm{tn}} \frac{\pi^2}{3} \left( \frac{T}{T_\trm{F}} \right)^2 \right]^{-1}
\end{equation}
for $T \gg T_\trm{F}$, where $\sigma_{zz}^\trm{a} (0)$ is the anomalous conductivity at zero temperature given by Eq.~\eqref{eq:LMC_approx}. Note that Eq.~\eqref{eq:high_T_asymp_appen} is equivalent to Eq.~\eqref{eq:high_T_asymp} in the main text.

\twocolumngrid

\end{document}